\documentclass[fleqn,usenatbib]{mnras}

\usepackage{newtxtext,newtxmath}
\usepackage{hyperref}
\usepackage[T1]{fontenc}

\DeclareRobustCommand{\VAN}[3]{#2}
\let\VANthebibliography\thebibliography
\def\thebibliography{\DeclareRobustCommand{\VAN}[3]{##3}\VANthebibliography}

\newcommand{\dd}[0]{\mathrm{d}}
\usepackage{graphicx}	% Including figure files
\usepackage{amsmath}	% Advanced maths commands
\title[]{Stellar/BH Population in AGN Disks: Direct Binary Formation from Capture Objects in Nuclei Clusters}

\author[]{
Yihan Wang$^{1,2}$\thanks{yihan.wang@unlv.edu},
Zhaohuan Zhu$^{1,2}$, and
Douglas N. C. Lin$^{3}$
\\
$^{1}$Nevada Center for Astrophysics, University of Nevada, Las Vegas, NV 89154, USA\\
$^{2}$Department of Physics and Astronomy, University of Nevada Las Vegas, Las Vegas, NV 89154, USA\\
$^{3}$Department of Astronomy \& Astrophysics, University of California, Santa Cruz, CA 95064, USA
}

\date{Accepted XXX. Received YYY; in original form ZZZ}

\pubyear{2015}

\begin{document}
\label{firstpage}
\pagerange{\pageref{firstpage}--\pageref{lastpage}}
\maketitle

\begin{abstract}
The Active Galatic Nuclei(AGN) disk has been proposed as a potential channel for the merger of binary black holes. The population of massive stars and black holes in AGN disks captured from the nuclei cluster plays a crucial role in determining the efficiency of binary formation and final merger rate within the AGN disks. In this paper, we investigate the capture process using analytical and numerical approaches. We discover a new constant integral of motion for one object's capture process. Applying this result to the whole population of the nuclei cluster captured by the AGN disk, we find that the population of captured objects depends on the angular density and eccentricity distribution of the nuclei clusters and is effectively independent of the radial density profile of the nuclei cluster and disk models. An isotropic nuclei cluster with thermal eccentricity distribution predicts a captured profile $\dd N/\dd r\propto r^{-1/4}$. The captured objects are found to be dynamically crowded within the disk. Direct binary formation right after the capture would be promising, especially for stars. The conventional migration traps that help pile up single objects in AGN disks for black hole mergers might not be required. 
\end{abstract}


%%%%%%%%%%%%%%%%%%%%%%%%%%%%%%%%%%%%%%%%%%%%%%%%%%

%%%%%%%%%%%%%%%%% BODY OF PAPER %%%%%%%%%%%%%%%%%%

\section{Introduction}

  Black hole (BH) mergers within the active galactic nucleus (AGN) disks have garnered considerable attention as a mechanism potentially explaining the existence of massive merging BHs\citep{Vokrouhlicky1998,Cuadra2009,McKernan2012,McKernan2014,Bartos2017,McKernan2018,Hoang2018,Secunda2019,Yang2019,Yang2019b,McKernan2020,Tagawa2020,Li2022,Bhaskar2022}. Nuclear stellar clusters (NSCs), as the densest environments of stars and BHs, coexist with most of the supermassive black holes (SMBHs)\citep{Paumard2006,Merritt2010,Genzel2010, kormendy2013}. For binaries around the SMBH, the typical binary semi-major axis is much larger than the critical size that gravitational wave radiation can drive the binary to merge within the Hubble time. However, the secular perturbations from the SMBH can excite the eccentricity of binaries to near unity, resulting in high gravitational wave radiation efficiency\citep{Wen2003,Antonini2015,Antonini2016,Stephan2016,VanLandingham2016,Petrovich2017,Liu2018,Hoang2018,Liu2019, Liu2019b,Bhaskar2022}. These eccentric binaries can shrink their semi-major axis by gravitational wave radiation much faster than circular binaries and could eventually merge within the Hubble time. However, due to the relatively large velocity dispersion and the deep gravitational potential of the SMBH, it is not easy to form binaries within the NSC. Unlike the galactic field with a binary fraction of approximately 50\%, the binary fraction in star clusters is believed to be less than 10\%\citep{Ivanova2005,Hurley2007,Sollima2007}. Moreover, most of the binaries in NSCs are located in the outskirts of the NSC, far away from the SMBH, where the secular perturbations from the SMBH are weak. The timescale for the SMBH to excite the eccentricity of binaries located at larger distance to the SMBH is very long. Therefore, a direct binary BH merger from the secular effects of the SMBH may not be efficient.

In the active phase of a galactic nucleus, gas is funneled into the nuclear region, leading to the assembly of accretion disks. The accretion process associated with high luminosity makes the SMBHs visible as active galactic nuclei. Once an AGN disk is formed, some of the stars and BHs in NSC may eventually be trapped within the disk through star/BH-disk interactions \citep{ Artymowicz1993,Rauch1995,Kennedy2016,Panamarev2018, Macleod2020,Fabj2020,davies2020,Generozov2023,Nasim2023}. The moving trapped stars/BHs excite density waves in the in AGN disk. The Lindblad resonance exerts torques on those embedded objects, pushing them to migrate within the disk \citep{Tanaka2002,Baruteau2010, Baruteau2011}. Previous studies \citep{Bellovary2016,Peng2021} have suggested that migration traps may be required to accumulate single BHs such that there's a dense region for binary to be formed through three-body/multi-body  scatterings \citep{Hills1975,Aarseth1976,Heggie1996,Leigh2018,Zevin2019} or gravitational wave captures\citep{OLeary2009,Kocsis2012,Gondan2018,Samsing2020,LiJ2022}. However, the hypothetical existence of the migration traps ignores multi-body interactions
during the migration process, stochastic diffusion due to chaotic torque in turbulent disks\citep{wu2024}, and relies on special idealized disk models.  

In addition to migration traps, \citet{Li2023,Rowan2022,Rozner2023,Delaurentiis2023} also investigated direct binary formation in AGN using gas dissipation with arbitrary initial BH populations. However, accurately estimating the rate of BH binary formation in AGN disks has proven challenging due to uncertainties in the density profile, initial mass function, binary fraction in NSC, and poorly constrained AGN disk models.

It is particularly challenging to accurately predict the BH binary merger rate within the AGN disk due to the unclear path from binary formation to the mergers. In the case of disk-embedded binaries, there are two primary mechanisms for reducing the semi-major axis to the regime of gravitational wave radiation: dynamical encounters \citep{Tagawa2021,Tagawa2021b,Wang2021,Samsing2022} and gas dissipation \citep{Baruteau2011,Li2022,LiR2023,Li2023b,Kaaz2023}.

Regarding dynamical encounters, if these embedded binaries exist in a denser dynamical environment, close interactions between these binaries and other single or binary systems can occur frequently. When the binaries are hard, characterized by significant environmental velocity dispersion, the scattering process statistically hardens the binary system. Notably, in this impulse regime, the scatterings tend to increase the eccentricity of the binaries, thereby accelerating the hardening process through enhanced gravitational wave radiation. However, this process is sensitive to the local populations of BHs and stars. The presence of hard binaries and an adequate number density are required for the hardening process to take effect.

In the case of gas-assisted mergers, the hardening process is more dependent on the binary orientation within the AGN disks. Retrograde binaries, with a greater relative velocity compared to the local disk velocity, efficiently dissipate orbital energy \citep{Li2023b}. However, for prograde binaries, there is no definitive conclusion on whether the gas can sufficiently reduce the binary's semi-major axis to reach the regime of gravitational wave radiation. Furthermore, the population of prograde/retrograde binaries, which relies on the initial conditions of binary formation, remains poorly constrained in current research.

All these subsequent steps (migration, binary formation, hardening) leading to BH merger depend on the capture process of stars and BHs by the AGN disks. {  When the relative velocity greatly exceeds the surface escape velocity, e.g. for stars on most orbits intersecting an AGN disk, the dominant force is aerodynamic drag. Conversely, when the relative velocity is considerably lower than the surface escape velocity of the object, e.g. for black holes and neutron stars on practically all orbits intersecting an AGN disk, the dominant force is gas dynamical friction.}
In this study, we investigate the capture {of stars and BHs from the nuclei cluster by} AGN disks, focusing on estimating the timescales associated with semi-major axis damping, inclination decrease, and eccentricity excitation/damping. By analyzing the dependence of these timescales on the initial orbital properties and their relative configuration with respect to the AGN disk, we aim to accurately predict the captured star/BH population in AGN disks.

Our paper is organized as follows. In Section 2, we investigate the disk-star/BH interactions, derive the equation of motion for this capture process, and obtain the timescales of semi-major axis, eccentricity, and inclination evolutions. In Section 3, we perform N-body simulations to test the physics picture we obtained from Section 2. In Section 4, based on the timescales and derived integral of motion, we calculate the population of the captured stars/BHs via Monte Carlo simulations and provide analytical expressions for stars/BHs population in AGN disks.

\section{Disk-star/BH interaction}\label{sec:calc}
\begin{figure}
	\includegraphics[width=1\columnwidth]{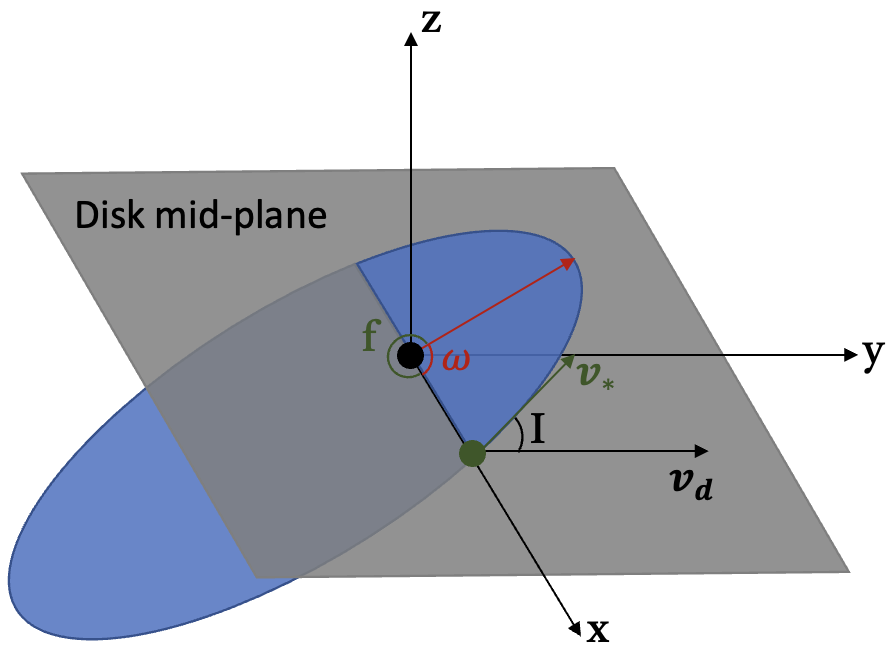}\\
    \caption{Schematics of the disk capture. The coordinates system is built in a way such that the ascending node of the star/CO orbit is on the x-axis.}
    \label{fig:schematics}
\end{figure}

Figure~\ref{fig:schematics} shows the schematics of the disk-star/BH interactions. The celestial body in orbit around the supermassive BH traverses the disk's mid-plane at two specific points known as the ascending node and descending node. Given the axis symmetry of the system, it is possible to establish Cartesian coordinates in such a way that the ascending node coincides with the positive x-axis. Consequently, the positional vector of the ascending and descending interacting nodes along the x-axis can be mathematically expressed as follows:
\begin{eqnarray}
r_\pm&=&\frac{p}{1\pm e\cos\omega}\\
\mathbf{r}&=&r(\pm1,0,0)
\end{eqnarray}
where {$\omega$ is the argument of periapsis}, $p=a(1-e^2)$ is the semi-latus rectum with semi-major axis $a$ and eccentricity $e$.  The corresponding velocities of the object ($\mathbf{v}_*$)and { the Keplerian rotating disk ($\mathbf{v}_d$)} at the crossing point are
\begin{eqnarray}
v_{*,\pm}&=&\sqrt{\frac{GM_{\rm tot}(1+e^2\pm 2e\cos\omega)}{p}}\\
v_{d,\pm}&=&\sqrt{\frac{GM_{\rm tot}}{r}}=\sqrt{\frac{GM_{\rm tot}(1\pm e\cos\omega)}{p}}\\
\mathbf{v}_{*,\pm} &=& \sqrt{\frac{GM_{\rm tot}}{p}}(-e\sin\omega, \pm\eta_\pm^2\cos I, \pm\eta_\pm^2\sin I)\\
\mathbf{v}_{d,\pm} &=& \sqrt{\frac{GM_{\rm tot}}{p}}(0,\pm\eta_\pm , 0)\\
\sigma_\pm &=& \sqrt{1+e^2\pm2e\cos\omega}\\
\eta_\pm &=& \sqrt{1\pm e\cos\omega}
\end{eqnarray}
where $M_{\rm tot}=M_{\rm SMBH}+m$ is the total mass of the supermassive black hole and crossing object, and $\omega$ is the argument of the periapsis. The relative velocity between the object and the disk can be expressed as
\begin{eqnarray}
\mathbf{v}_{\rm rel}&=&\sqrt{\frac{GM_{\rm tot}}{p}}\bigg(- e\sin\omega,\pm\eta_\pm^2\cos I \mp\eta_\pm,\pm\eta_\pm^2\sin I\bigg) 
\end{eqnarray}
And the magnitude of the relative velocity is
\begin{eqnarray}
v^2_{\rm rel} &=& \frac{GM_{\rm tot}}{p}(\sigma_\pm^2+\eta_\pm^2-2\eta_\pm^3\cos I)\nonumber\\
&=&\frac{2GM_{\rm tot}\eta_\pm^3}{p}\left(\frac{\sigma_\pm}{\eta_\pm^2}-\cos I\right)
\end{eqnarray}
where we use 
\begin{eqnarray}
\frac{\sigma_\pm^2+\eta_\pm^2}{2\sigma_\pm\eta_\pm}=1+\frac{e^2\cos^2\omega}{8}+\frac{e^3(1-e^2)\cos\omega}{8(1+e^2)^{3/2}}+\mathcal{O}(e^4)+\mathcal{O}(\cos^3\omega)\sim1
\end{eqnarray}

Upon the celestial object crosses the disk, it becomes subject to a drag force. The magnitude of this force depends on the ratio between the relative velocity of the object with respect to the disk and the surface escape velocity of the object. Two distinct mechanisms can dominate the drag force: aerodynamic drag force and gas dynamical friction.

When the relative velocity greatly exceeds the surface escape velocity of the object, the dominant force is the aerodynamic drag force, which is directly proportional to the square of the relative velocity ($v_{\rm rel}^2$). Conversely, when the relative velocity is considerably lower than the surface escape velocity of the object, the dominant force is gas dynamical friction, which is inversely proportional to the square of the relative velocity ($v_{\rm rel}^{-2}$) \citep{Ostriker1999}.

If the aerodynamic drag force dominates the overall drag force, it can be described by the following equation:
\begin{equation}
    \mathbf{F}_{\rm aero} = -\pi R^2 \rho_g {v}_{\rm rel} \mathbf{v}_{\rm rel}\label{eq:f_aero}
\end{equation}
where $\rho_g$ is the gas density of the disk, $\mathbf{v}_{\rm rel}$ is the relative velocity between the crossing object and the disk, and $R$ is the effective radius of the crossing object. The effective radius is defined as the maximum value between the physical radius and gravitational radius $Gm/v_{\rm rel}^2$.

If the dynamical friction dominates the overall drag force, the drag force can be described as,
\begin{equation}
    \mathbf{F}_{\rm dyn} = -\mathcal{I}\frac{4\pi(Gm)^2\rho_g}{v^3_{\rm rel}}  \mathbf{v}_{\rm rel}\label{eq:f_dyn}
\end{equation}
where $\mathcal{I}$ is a function of Mach number $\mathcal{M}$.
\begin{eqnarray*}
\mathcal{I}=\left\{
\begin{aligned}
    \frac{1}{2}\ln\bigg(\frac{1+\mathcal{M}}{1-\mathcal{M}}\bigg)-\mathcal{M}, \mathcal{M}<1\\
    \frac{1}{2}\ln\bigg(1-\frac{1}{\mathcal{M}^2}\bigg)+\Lambda,\mathcal{M}>1
\end{aligned}
\right.
\end{eqnarray*}
with Coulomb logarithm $\Lambda = \log(R_{\rm max}/R_{\rm eff})$, where $R_{\rm max}$ is the typical size of the medium. In the subsonic regime where $\mathcal{M}<1$, $\mathbf{F}_{\rm dyn}$ is asymptotic to $\mathcal{M}/3$ and in the supersonic regime where $\mathcal{M}>1$, $\mathbf{F}_{\rm dyn}$ is asymptotic to $\mathcal{M}^{-2}$\footnote{Note that in the subsonic regime, the gas dynamical friction is typically much smaller than the dynamical friction in collisionless medium, e.g. background stars, where
$I_{\rm collisionless}=\Lambda[{\rm erf}(X) - \frac{2X}{\sqrt{\pi}}e^{-X^2}]$
where $X={v_{\rm rel}}/{\sigma\sqrt{2}}$ with velocity dispersion of the collisionless medium $\sigma$.}.

{The criteria remain essentially the same for cases where the gravitational radius dominates over the physical radius or when gas dynamical friction dominates over aerodynamic drag (considering the physical radius), where
\begin{eqnarray}
\frac{v_{\rm rel}^2}{v_{\rm esc}^2} < 1 .
\end{eqnarray}
}

For main sequence stars, BHs, and neutron stars in circular orbits, the criteria writes as,
\begin{eqnarray}
1-\cos I_* &<& 2\times10^{-3} \left(\frac{m}{1M_\odot}\right)\left(\frac{M_{\rm SMBH}}{10^8M_\odot}\right)^{-1}\left(\frac{p}{\rm 0.01 pc}\right)\left(\frac{R}{1 R_\odot}\right)^{-1}\\
1-\cos I_{\rm BH} &<& 10^3 \left(\frac{M_{\rm SMBH}}{10^8M_\odot}\right)^{-1}\left(\frac{p}{\rm 0.01 pc}\right)\\
1-\cos I_{\rm NS} &<& 2\times10^2 \left(\frac{M_{\rm SMBH}}{10^8M_\odot}\right)^{-1}\left(\frac{p}{\rm 0.01 pc}\right)
\end{eqnarray}
For compact objects, such as black holes or neutron stars, the gravitational radius generally dominates over the physical radius. In addition, gas dynamical friction tends to dominate over aerodynamic drag when considering the physical radius ({for BHs, the event horizon}).

On the other hand, for main sequence stars, the physical radius typically outweighs the gravitational radius in the majority of parameter space, except in cases where the inclination is sufficiently low. Consequently, there exists a critical inclination for main sequence stars, below which the gravitational radius becomes significant compared to the physical radius, and gas dynamical friction becomes dominant over aerodynamic drag,
\begin{equation}
I_{\rm d}\sim 0.05\left(\frac{m}{1M_\odot}\right)^{1/2}\left(\frac{M_{\rm SMBH}}{10^8M_\odot}\right)^{-1/2}\left(\frac{p}{\rm 0.01 pc}\right)^{1/2}\left(\frac{R}{1 R_\odot}\right)^{-1/2}\label{eq:Ic}
\end{equation}

Also see \citep{Grishin2015} for a comparison between gas drag and gas dynamical friction for planetesimals moving in protoplanet disk.

\subsection{Aerodynamic drag}\label{sec:aero}
\begin{figure}
	\includegraphics[width=\columnwidth]{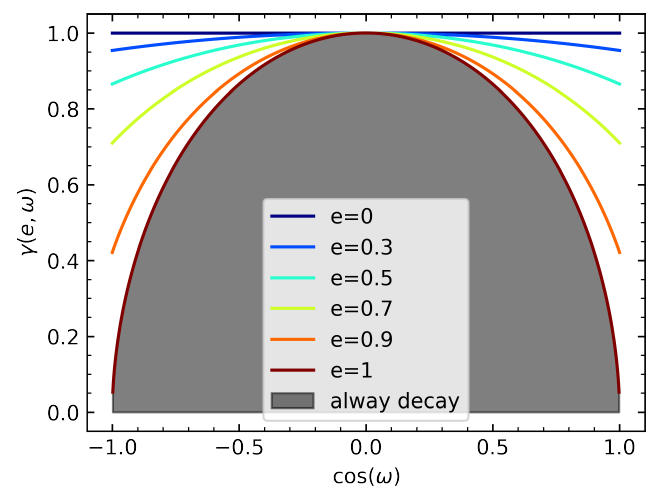}\\
    \caption{$\gamma(e,\omega)$ as a function of $\cos\omega$ for different eccentricities. $\cos I < \gamma$ indicates angular momentum damping.}
    \label{fig:gamma}
\end{figure}

If the disk is assumed to be thin (h$\ll$1), and the drag force is dominant by the aerodynamic drag with geometry cross-section, the specific momentum and specific angular momentum change per disk cross are,
\begin{eqnarray}
    \Delta\mathbf{v}&=&\frac{\mathbf{F}}{m}\Delta t
    \sim-\lambda \mathbf{v}_{\rm rel}\\
    \Delta \mathbf{L}&=&\mathbf{r}\times \Delta\mathbf{v}=-\lambda \mathbf{r}\times \mathbf{v}_{\rm rel}\label{eq:dL}\\
    \Delta E &=& \frac{\mathbf{F}}{m} \cdot \mathbf{v}_{*,\pm}\Delta t=-\lambda\mathbf{v}_{\rm rel}\cdot \mathbf{v}_{*,\pm} 
\end{eqnarray}
where $\Delta t \sim H/v_{\rm rel,\perp}$ and $\lambda =\frac{\pi R_{*}^2\Sigma}{m}\frac{v_{\rm rel}}{v_{\rm rel,z}}$. The specific energy, specific angular momentum, and the relative velocity are,
\begin{eqnarray}
E &=& \frac{GM_{\rm tot}}{-2a}\\
\mathbf{L}&=&L(0,-\sin I,\cos I)\label{eq:L}
\end{eqnarray}
where $L = \sqrt{GM_{\rm tot}p}$ and $I$ is the orbit inclination. Based on Equation~\ref{eq:dL}, the specific angular momentum and energy change per disk cross are,
\begin{eqnarray}
\Delta \mathbf{L}&=& \lambda L(0,\sin I,\frac{1}{\eta_\pm}-\cos I)\\
\Delta E &=&\lambda\frac{GM_{\rm tot}}{p}(\eta_\pm^3\cos I-\sigma_\pm^2) 
\end{eqnarray}
Assuming that the relative angular momentum change per disk crossing, denoted as $\lambda$, is small ($\lambda \ll 1$), we can use Equation~\ref{eq:L} and the relationship $\cos I = L_z 
/ L$ to derive the linear equation of motion for the angular momentum and inclination,
\begin{eqnarray}
\frac{\dd L^2}{L^2\dd(t/T)} &=& 4\sqrt{2}\alpha\frac{\Sigma}{\Sigma_*}\frac{\cos I - \gamma}{\sin I}\sqrt{ \delta -\cos I} + \mathcal{O}(\lambda^2)\label{eq:dLdt}\\
\frac{\dd \cos I}{\dd(t/T)} &=& 2\sqrt{2}\alpha\frac{\Sigma}{\Sigma_*}{\sin I}\sqrt{ {\delta}-\cos I} + \mathcal{O}(\lambda^2)\label{eq:dIdt}\\
 \frac{\dd L^2}{2L^2\dd\cos I}&=&\frac{\cos I-\gamma}{\sin^2 I}\label{eq:dLdI}\\
 \frac{\dd E}{E\dd(t/T)} &=& \frac{4\sqrt{2}}{1-e^2}\bar{\alpha}\frac{\Sigma}{\Sigma_*}\frac{\bar{\gamma}-\cos I}{\sin I}\sqrt{ \delta -\cos I} 
\end{eqnarray}
where 
\begin{eqnarray}
%\lambda &=&\frac{\Sigma}{\Sigma_*}\sqrt{\frac{2}{\eta_\pm}}\frac{\sqrt{\sigma_\pm/\eta_\pm^2-\cos I}}{\sin I}  \\
{\delta}&=&\frac{1}{2}(\frac{\sigma_+}{\eta_+^2} + \frac{\sigma_-}{\eta_-^2} )\\
\alpha &=&  \frac{1}{2}(\sqrt{\frac{1}{\eta_+^3}} +\sqrt{\frac{1}{\eta_-^3}}) \\
\beta &=& \frac{1}{2}(\sqrt{\frac{1}{\eta_+}} + \sqrt{\frac{1}{\eta_-}}) \\
\gamma &=& \beta/\alpha\\
\bar{\alpha} &=&  \frac{1}{2}(\sqrt{\eta_+^5} +\sqrt{\eta_-^5}) \\
\bar{\beta} &=& \frac{1}{2}(\sqrt{\frac{\sigma_+^4}{\eta_+}} + \sqrt{\frac{\sigma_-^4}{\eta_-}}) \\
\bar{\gamma} &=& \bar{\beta}/\bar{\alpha}
\end{eqnarray}
$T=2\pi\sqrt{a^3/(GM_{\rm tot})}$ is the period of the crossing orbit and $\Sigma_*=\frac{m}{\pi R_*^2}$ is the surface density of the crossing object. All these parameters are defined to be equal to unity if $e=0$.  Using the relation
\begin{eqnarray}
\frac{\dd a}{a} &=& -\frac{\dd E}{E}\\
\frac{\dd e}{e} &=& \frac{1-e^2}{2e^2}\bigg(\frac{\dd a}{a}-\frac{\dd p}{p}\bigg)
\end{eqnarray}

We can also get
\begin{eqnarray}
\frac{\dd a}{a\dd(t/T)} &=& \frac{4\sqrt{2}}{1-e^2}\bar{\alpha}\frac{\Sigma}{\Sigma_*}\frac{\cos I-\bar{\gamma}}{\sin I}\sqrt{ \delta -\cos I} \label{eq:dadt}\\
\frac{\dd e}{e\dd(t/T)} &=& \frac{1}{2e^2}\frac{\Sigma}{\Sigma_*}\frac{4\sqrt{2}}{\sin I}\sqrt{ \delta -\cos I}\bigg((\bar{\alpha}\cos I - \bar{\beta})\nonumber\\
&-& (1-e^2)(\alpha\cos I - \beta)\bigg)\label{eq:dedt}
\end{eqnarray}
$\Sigma$ at the two crossing points $r_\pm$ is assumed to be nearly constant. This assumption holds true when $e\cos\omega$ is not close to unity, indicating that the orbit is not extremely eccentric with $\cos\omega \approx \pm 1$. If $e\cos\omega$ is close to unity, it implies that $r_-$ is orders of magnitude smaller than $r_+$, causing the surface density $\Sigma(r_-)$ to be much larger than $\Sigma(r_+)$. Consequently, the dissipation is primarily influenced by the crossing point $r_-$, and terms involving $\alpha$, $\beta$, and $\gamma$ with $\sigma_+$ and $\eta_+$ can be disregarded.

\subsubsection{Inclination damping and eccentricity evolution}
\begin{figure}
	\includegraphics[width=\columnwidth]{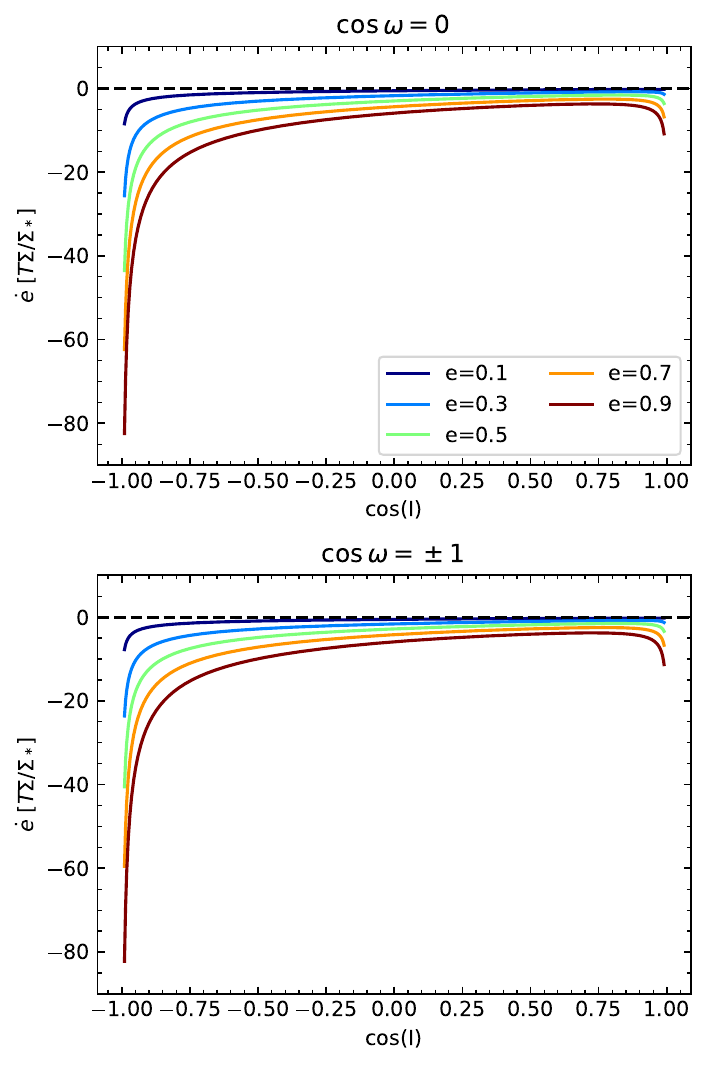}\\
    \caption{$\dot{e}$ as a function of inclination for different $e$. The upper panel shows the case with $\cos\omega=0$ where the two crossing points have the same distance to the SMBH. The bottom panel shows the case with $\cos\omega=\pm1$ where the periapsis and apoapsis lie within the disk. Eccentricity will always decrease for all inclinations. }
    \label{fig:e-aero}
\end{figure}

Equation~\ref{eq:dIdt} indicates that $\dd \cos I / \dd t$ is always greater than zero, implying that the inclination evolves towards $\cos I \rightarrow 1$, or in other words, the inclination tends to align with the disk plane, which is consistent with direct 3-D hydrodynamical simulations \citep{Rein2012,Arzamasskiy2018,Zhu2019}. For objects in nuclei clusters with inclined orbits, the aerodynamic drag will gradually align the orbits with the disk. If the time is long enough, all objects with different orbital configurations will be captured by the disk in prograde orbits. However, in reality, the lifetime of the AGN disk might be short compared to the capture timescale. Therefore, only a fraction of the objects in the nuclei cluster can be captured.

In Equation~\ref{eq:dLdt}, we have $\dd L^2/\dd t \propto \cos I - \gamma$. Therefore, if $\cos I > \gamma$, $\dd L^2/\dd t$ will be positive, indicating the angular momentum will increase, otherwise, it will decrease. From Figure~\ref{fig:gamma}, we can observe that $\gamma$ is always smaller than one. The criterion for angular momentum damping is $\cos I < \gamma$. If the inclination is low enough and the periapsis/apoapsis are closer to the disk mid-plane ( $\cos\omega\rightarrow\pm1$), the angular momentum can increase. In this configuration, at the apoapsis, the local disk Keplerian velocity is much higher than the velocity of the orbiter. Thus, the disk can accelerate the orbiter and transfer angular momentum to the orbiter. From Equation~\ref{eq:dadt}, due to $\bar{\alpha}>0$ and $\bar{\gamma}>1$, the semi-major axis always decays. Therefore, the angular momentum increase indicates fast eccentricity damping.

 Note that this angular momentum increase criterion also requires the aerodynamic drag to dominate over the gas dynamical friction. Therefore, the inclination must be larger than the critical inclination indicated by Equation~\ref{eq:Ic}. Therefore, the parameter space for angular momentum increase also requires $\cos I_{\rm d} > \gamma$. Otherwise, the required inclination for angular momentum increase is below the critical inclination where aerodynamic drag dominates the total drag force.

Regarding the eccentricity evolution, one can prove that Equation~\ref{eq:dedt} is always negative if $e\neq0$. So the eccentricity will always decrease. Figure~\ref{fig:e-aero} shows the value of $\dot{e}$ as a function of $\cos I$ for different $e$ and $\omega$. As shown in the upper panel and bottom panel, for $\cos\omega=0$, where the two crossing points have the same distance to the SMBH, or for $\cos\omega=\pm1$, where the periapsis and apoapsis lie within the disk, $\dot{e}$ is always negative for various initial eccentricities and inclinations. Therefore, the eccentricity will be damped during the capture process.

\subsubsection{Timescales}

 From Equation~\ref{eq:dLdt} to \ref{eq:dedt}, we can obtain the timescales for semi-lotus rectum, semi-major axis, inclination and eccentricity evolution,
\begin{eqnarray}
\tau_{\rm p, aero} &=& \bigg|\frac{p}{\dot{p}}\bigg|=\bigg|\frac{L^2}{\dot{L^2}}\bigg|= \frac{\sin I}{4\sqrt{2}\alpha|\cos I - \gamma|\sqrt{{\delta}-\cos I}}\frac{\Sigma_*}{\Sigma}T\\
\tau_{\rm a, aero} &= &\bigg|\frac{a}{\dot{a}}\bigg|= \frac{(1-e^2)\sin I}{4\sqrt{2}\bar{\alpha}(\bar{\gamma}-\cos I)\sqrt{{\delta}-\cos I}}\frac{\Sigma_*}{\Sigma}T\label{eq:t_a_aero}\\
\tau_{\rm I, aero} &=& \bigg|\frac{I\sin I}{\dot{\cos I}}\bigg| = \frac{I}{2\sqrt{2}\alpha\sqrt{{\delta}-\cos I}}\frac{\Sigma_*}{\Sigma}T\label{eq:t_inc,aero}\\
\tau_{\rm e, aero} &=& \frac{2e^2}{(1-e^2)}\left|\frac{1}{\tau_{\rm a, aero}}-\frac{1}{\tau_{\rm p, aero}}\right|^{-1}
\end{eqnarray}
For the aerodynamic drag-dominated regime, the timescales for $a$, $e$ and $I$ evolution are at the order of magnitude of $\frac{\Sigma_*}{\Sigma}T$. For main-sequence stars, if we use the mass-radius relation
\begin{eqnarray}
R = R_\odot(\frac{m}{M_\odot})^{3/4}
\end{eqnarray}
we can get
\begin{eqnarray}
\Sigma_* = \Sigma_\odot \left(\frac{m}{M_\odot}\right)^{-1/2}=1.39\times10^{11}\left(\frac{m}{M_\odot}\right)^{-1/2} \rm g/cm^2
\end{eqnarray}

So, for massive stars, the timescales for semi-major axis, eccentricity, and inclination evolution are shorter.

\begin{figure}
	\includegraphics[width=\columnwidth]{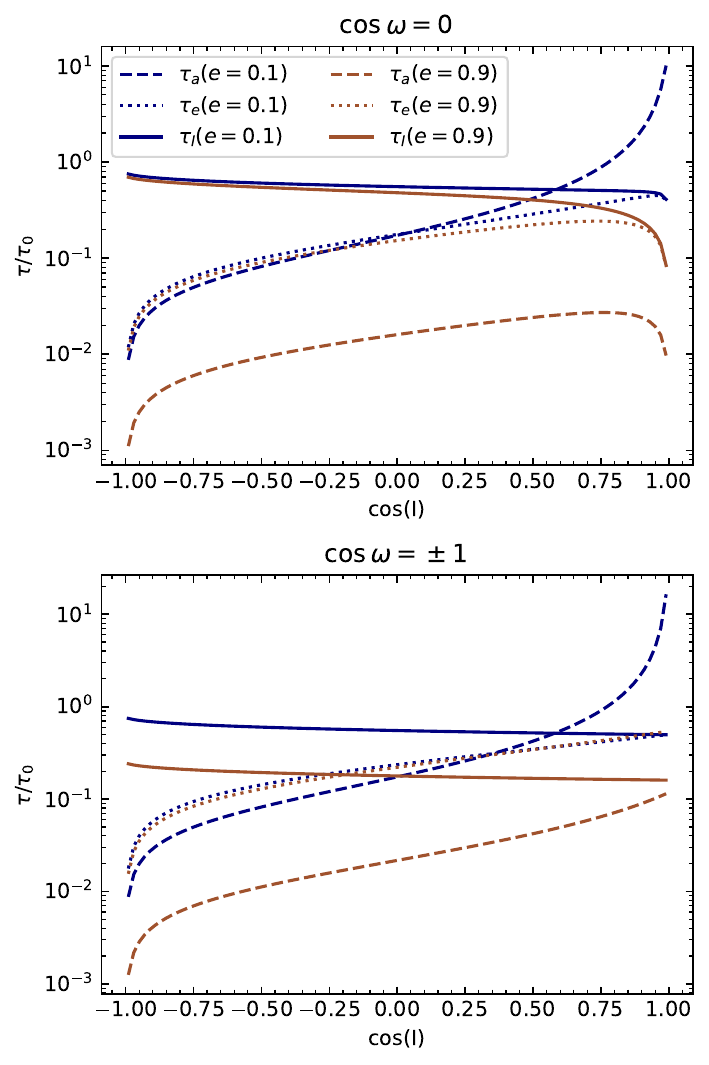}\\
    \caption{Timescales of semi-major axis, eccentricity and inclination evolution in a unit of $\tau_0=T\Sigma_*/\Sigma$ for stars. The upper panel shows the case for $\cos\omega=0$ where the two crossing points have the same distance to the SMBH. The bottom panel shows the case with $\cos\omega=\pm1$ where the periapsis and apoapsis lie within the disk. In general, the semi-major damping timescale is much shorter than the inclination damping timescale and the inclination damping timescale is shorter than the eccentricity excitation/damping timescale. }
    \label{fig:tau-aero}
\end{figure}
Figure~\ref{fig:tau-aero} shows the timescales of semi-major axis, eccentricity, and inclination evolution in a unit of $\tau_0=T\frac{\Sigma_*}{\Sigma}$ for different $e$ and $\omega$. The upper panel shows the case $\cos\omega=0$, where the two crossing points on disk mid-plane have the same distance to the central SMBH. For low  eccentricity, we can see the timescale for semi-major axis damping is much shorter than the timescale of inclination damping with high inclinations and is much longer than the timescale of inclination damping with low inclinations. For high eccentricities, the semi-major axis damping timescale is always shorter than the other two timescales. The timescale for eccentricity is always shorter than the inclination damping timescale for both low and high eccentricities. Therefore, for high-inclination orbits, the sem-major axis and eccentricity decrease fast, with nearly unchanged inclination. As the inclination decreases and the orbit becomes circularized, the semi-major axis damping timescale becomes longer and longer, thus the orbit will eventually stall at a certain semi-major axis.

The bottom panel shows another case, where the periapsis and apoapsis lie on the disk mid-plane, the general picture is very similar to the case with $\cos\omega=0$, except for the inclination damping timescale is more sensitive to the eccentricity. 

\subsubsection{Integral of motion}\label{sec:c of motion}

As indicated by Figure~\ref{fig:gamma}, the $\gamma$ is nearly constant within one order of magnitude and is close to unity for almost all eccentricities and $\omega$. Therefore, we can integrate Equation~\ref{eq:dLdI} to be
\begin{eqnarray}
L\cos^2(I/2)\cot^{\gamma-1}(I/2)=\rm const \label{eq:LI}
\end{eqnarray}
with approximation $\gamma\sim1$, this expression can be further simplified to,
\begin{eqnarray}
L\cos^2(I/2)=\rm const
\end{eqnarray}
From the previous discussion we learn that if the capture timescale $\tau_{\rm I, aero}$ is shorter than the lifetime of the disk, the objects in the nuclei cluster will be eventually captured by the disk. As a result, the final inclination ($I_{\rm f}$) will be effectively zero. The final captured angular momentum can be calculated by
 \begin{equation}
 L_{\rm f} = \sqrt{GM_{\rm tot}a_{\rm f}(1-e_f^2)} = \sqrt{GM_{\rm tot}a_0(1-e_0^2)}\cos^2(I_0/2)
 \end{equation}
From the previous discussion, we know that as the inclination is below the critical inclination $I_{\rm d}$, the dynamical friction takes over the aerodynamical drag. Later, we will show that for gas dynamic friction, this integral of motion holds almost true as well, and the eccentricity will always be circularized once the inclination is low enough. Therefore, the captured objects will effectively have zero eccentricity. In this way, the final captured semi-major axis can be well described as
\begin{eqnarray}
a_{\rm f} = a_0(1-e^2_0)\cos^4(I_0/2)
\end{eqnarray}

This integral of motion is effectively independent of $\lambda$, and thus is independent of the surface density of the disk. Since $M_{\rm tot}=M_{\rm SMBH}+m\sim M_{\rm SMBH}$, this constant is also essentially independent of the mass of the crossing objects. Therefore, even if the crossing objects are accreting during the capture process, the integral of motion still holds true. The captured objects will shrink the semi-major axis by a factor of $\frac{1}{(1-e_0^2)\cos^4(I_0/2)}$. Therefore, the initial retrograde obiters with high eccentricity will be captured in the innermost region of the disk. For orbiters with an initial inclination larger than the critical inclination
\begin{eqnarray}
I_{\rm TDE} = 2\arccos\left(\left(\frac{R_{\rm TDE}}{a_0(1-e_0^2)}\right)^{1/4}\right)\sim \pi-2\left(\frac{R_{\rm TDE}}{a_0(1-e_0^2)}\right)^{1/4}
\end{eqnarray}
where $R_{\rm TDE}$ is the tidal disruption radius of the SMBH on orbits, they will be tidally disrupted by the SMBH.

There's also another critical angle $I_c$, if the initial inclination is greater than this angle, the stars will be tidally disrupted in a retrograde orbit out of the disk. Following the integral of motion, we can obtain the critical angle
\begin{eqnarray}
a_0(1-e_0^2)\cos^4(I_c/2)&\sim&R_{\rm TDE}\cos^4(\pi/4)\\
I_c&=&2\arccos\left(\frac{\sqrt{2}}{2}\left(\frac{R_{\rm TDE}}{a_0(1-e^2_0)}\right)^{1/4}\right)\nonumber\\
&\sim& \pi - \sqrt{2}\left(\frac{R_{\rm TDE}}{a_0(1-e_0^2)}\right)^{1/4}
\end{eqnarray}

\begin{figure*}
    \includegraphics[width=2\columnwidth]{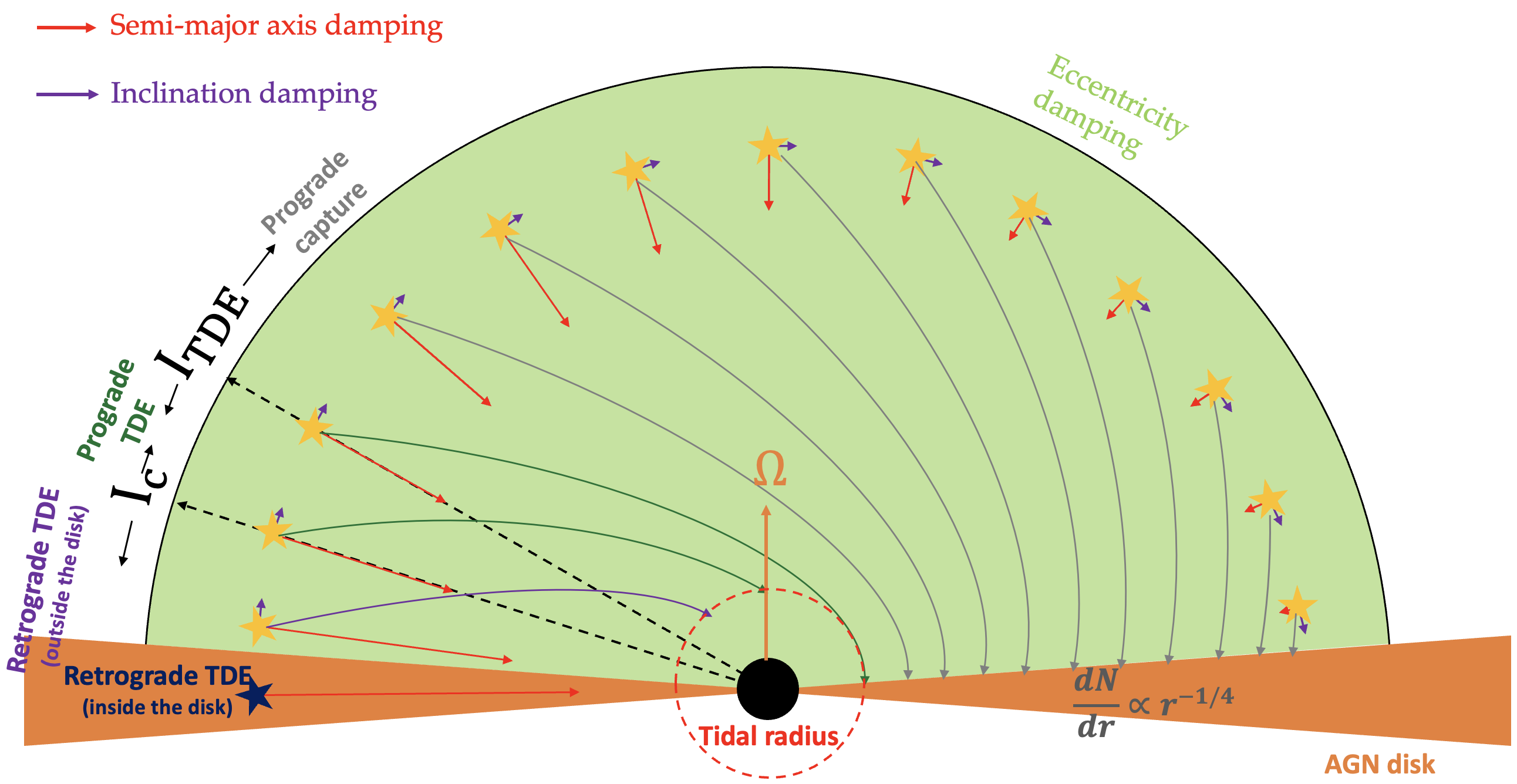}\\
    \caption{Carton shows the capture of stars with different initial inclinations at a given fixed semi-major axis. High-inclination stars will end up with smaller semi-major axis while low-inclination stars tend to have relatively larger final semi-major axis. Orbits with inclination $I$ smaller than $I_{\rm TDE}$ will eventually be captured by the AGN disk while orbits with inclination $I$ greater than $I_{\rm TDE}$ will be tidally disrupted by the SMBH. The trajectories and critical angles are not to scale, especially the semi-major axis that shrinks fast than the inclination damping and eccentricity evolution.}
    \label{fig:carton-star}
\end{figure*}

The general picture for the stellar capture is shown in Figure~\ref{fig:carton-star},  where for a given semi-major axis, two critical angles $I_c$ and $I_{\rm TDE}$ divide the parameter space into two different regimes. For very high inclination orbits, the semi-major axis shrinks very fast by a factor of $\frac{1}{(1-e_0^2)\cos^4(I_0/2)}$. This contributes to the TDE by disk captures. For retrograde orbits not associated with extremely high inclination, their semi-major axis damping would be slower and associated with eccentricity decrease. 

\subsection{Gas dynamical friction}
If the dynamical radius $G^2m^2/v_{\rm rel}^2$ is larger than the geometry radius $R_*$ or the gas dynamical friction dominant over the aerodynamic drag, the angular momentum equation should be
\begin{eqnarray}
    \Delta \mathbf{L}&=&-\lambda^\prime \mathbf{r}\times \mathbf{v}_{\rm rel}\\
    \Delta E &=& -\lambda^\prime\mathbf{v}_{\rm rel}\cdot \mathbf{v}_{*,\pm} 
\end{eqnarray}
where
{ 
\begin{equation}
\lambda^\prime =\frac{\pi G^2m\Sigma}{v^4_{\rm rel}}\frac{v_{\rm rel}}{v_{\rm rel,z}} = \frac{\pi m\Sigma p^2}{M^2_{\rm tot}} \frac{1}{ 2\sqrt{2}\eta_\pm^{13/2}\sin I (\frac{\sigma_\pm}{\eta_\pm^2}-\cos I)^{3/2}}
\end{equation}
}

 Similar to Section~\ref{sec:aero}, we can write the equation of motion of the angular momentum, inclination, semi-major axis and eccentricity,

{ 
\begin{eqnarray}
\frac{\dd L^2}{L^2\dd(t/T)} &=&\sqrt{2}\kappa\frac{\pi m\Sigma p^2}{M^2_{\rm tot}}\frac{\cos I - \zeta}{\sin I (\delta -\cos I)^{3/2}}\label{eq:dLdt2}\\
\frac{\dd \cos I}{\dd(t/T)} &=&\frac{\sqrt{2}}{2}\kappa\frac{\pi m\Sigma p^2}{M^2_{\rm tot}}\frac{\sin I}{(\delta -\cos I)^{3/2}} \label{eq:dIdt2}\\
\frac{\dd a}{a \dd(t/T)} &=&\frac{\sqrt{2}}{1-e^2}\bar{\kappa}\frac{\pi m\Sigma p^2}{M^2_{\rm tot}}\frac{\cos I - \bar{\zeta}}{\sin I(\delta -\cos I)^{3/2}}\label{eq:dadt2}\\
\frac{\dd e}{e \dd(t/T)} &=&\frac{1}{\sqrt{2}e^2}\frac{\pi m\Sigma p^2}{M^2_{\rm tot}}\frac{1}{\sin I(\delta -\cos I)^{3/2}}\nonumber\\&\times&\bigg((\bar{\kappa}\cos I -\bar{\xi}) - (1-e^2)(\kappa\cos I -\xi)\bigg)\\
\frac{\dd L^2}{2L^2\dd\cos I}&=&\frac{\cos I-\zeta}{\sin^2 I}\label{eq:dLdI2}
\end{eqnarray}
}
\begin{eqnarray}
\kappa &=&  \frac{1}{2}(\sqrt{\frac{1}{\eta_+^{15}}} + \sqrt{\frac{1}{\eta_-^{15}}}) \\
\xi &=& \frac{1}{2}(\sqrt{\frac{1}{\eta_+^{13}}} + \sqrt{\frac{1}{\eta_-^{13}}}) \\
\zeta &=& \frac{\xi}{\kappa}\\
\bar{\kappa} &=&  \frac{1}{2}(\sqrt{\frac{1}{\eta_+^{7}}} + \sqrt{\frac{1}{\eta_-^{7}}}) \\
\bar{\xi} &=& \frac{1}{2}(\sqrt{\frac{\sigma_+^4}{\eta_+^{13}}} + \sqrt{\frac{\sigma_-^4}{\eta_-^{13}}}) \\
\bar{\zeta} &=& \frac{\bar{\xi}}{\bar{\kappa}}
\end{eqnarray}

\subsubsection{Capture, eccentricity damping, and excitation}
Equation~\ref{eq:dIdt2} indicates that $\dd\cos I/\dd t$ is always positive. Therefore, similar to aerodynamic drag, the gas dynamical friction decreases the inclination of the orbiters. For Equation~\ref{eq:dLdt2}, if $\cos I-\xi$ is negative, the angular momentum will decrease while if $\cos I-\xi$ is positive, the angular momentum will increase. The reason for the angular momentum increase/decrease is the same as the aerodynamic drag that has been discussed.
\begin{figure}
	\includegraphics[width=\columnwidth]{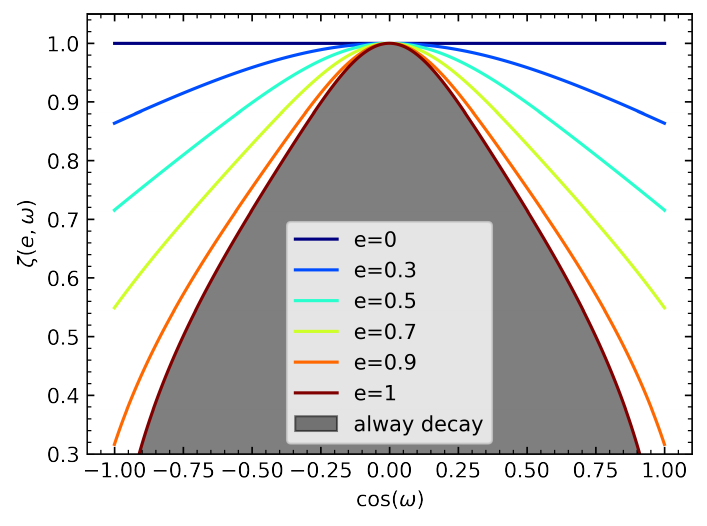}\\
    \caption{ $\zeta(e,\omega)$ as a function of $\cos\omega$ for different eccentricities. $\cos I < \zeta$ indicates angular momentum damping.}
    \label{fig:zeta}
\end{figure}
Figure~\ref{fig:zeta} shows $\zeta$ as functions of $e$ and $\omega$. Similar to Figure~\ref{fig:gamma}, $\zeta$ is close to unity but exhibits a more pronounced curvature as $e$ increases. 

For gas dynamical friction, the criterion for eccentricity damping is
\begin{eqnarray}
(\bar{\kappa}\cos I - \bar{\xi})-(1-e^2)(\kappa\cos I - \xi)<0\label{eq:e-c-dyn}
\end{eqnarray}
It's eas to calculate the critical inclination $I_{\rm e}$ for BH eccentricity damping,
\begin{eqnarray}
\cos I_{\rm e} = \frac{\bar{\xi}-(1-e^2)\xi}{\bar{\kappa}-(1-e^2)\kappa}\label{eq:ie}
\end{eqnarray}

\begin{figure}
	\includegraphics[width=\columnwidth]{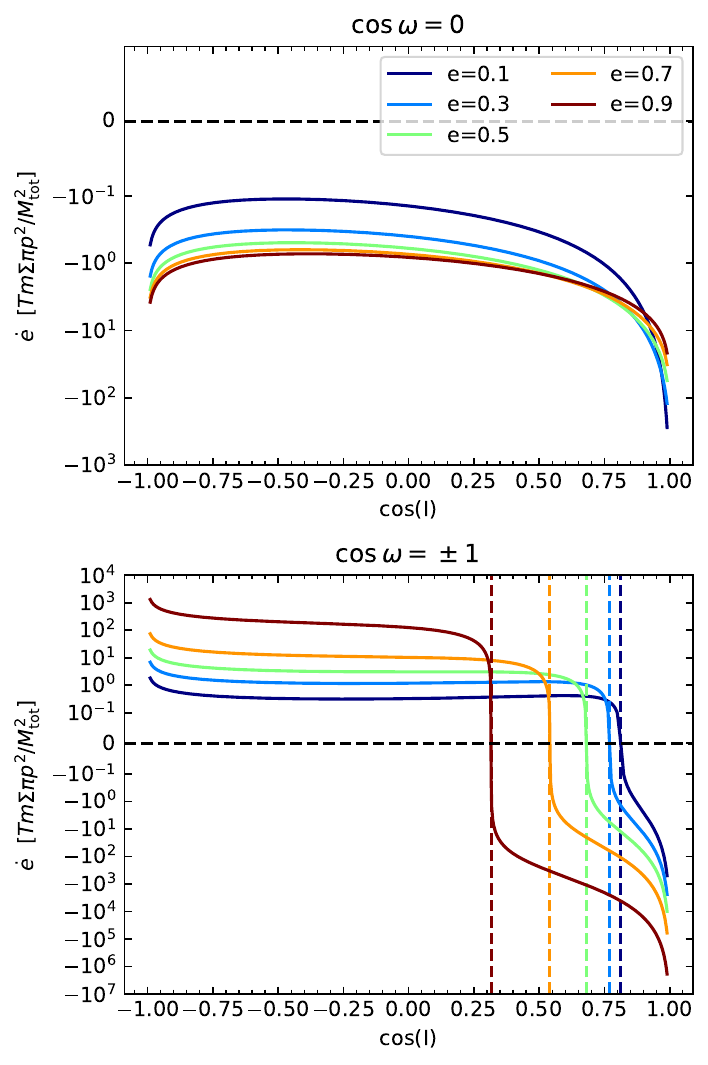}\\
    \caption{Similar to Figure~\ref{fig:e-aero}, but for gas dynamical friction. Unlike the aerodynamical drag, there is parameter space for eccentricity excitation. The vertical dashed lines show the critical inclination for eccentricity excitation/damping indicated by Equation~\ref{eq:ie}.}
    \label{fig:e-dyn}
\end{figure}
Figure~\ref{fig:e-dyn} illustrates the $\dot{e}$ as a function of inclination for different $e$ and $\omega$. Similar to Figure~\ref{fig:e-aero}, the upper panel/bottom panel shows the case for $\cos\omega=0$ and $\pm1$, respectively. For $\cos\omega=0$, similar to the aerodynamical drag, the eccentricity always decays, even if the orbit is purely retrograde. As $\cos I_{\rm e}$ becomes larger than -1, retrograde orbits with inclination $\cos I<\cos I_{\rm e}$ will undergo eccentricity excitation. For eccentricity damping to appear, we need  $\cos I_{\rm e} > -1$, such that any inclination $I<I_{\rm e}$ will undergo eccentricity damping. The criterion for $\cos I_{\rm e} > -1$ is $|\cos\omega|\sim>0.1$ for various eccentricities. Therefore, there's only a small region in the parameter space where eccentricity damping never appears. 

\subsubsection{Timescales}
%\begin{figure*}
%    \includegraphics[width=2\columnwidth]{carton-bh.png}\\
%    \caption{Similar to Figure~\ref{fig:carton-star}, but for BHs.}
%    \label{fig:carton-bh}
%\end{figure*}

Similar to Section~\ref{sec:aero}, we can get the  semi-latus rectum, inclination, semi-major axis, and eccentricity evolution timescales,
{ 
\begin{eqnarray}
\tau_{\rm p, dyn} &=& \frac{\sin I(\delta-\cos I)^{3/2}}{\sqrt{2}\kappa|\cos I -\zeta|}\frac{M_{\rm tot}}{m}\frac{M_{\rm tot}}{\Sigma \pi p^2} T\\
\tau_{\rm I, dyn} &=& \frac{\sqrt{2}I(\delta-\cos I)^{3/2}}{\kappa}\frac{M_{\rm tot}}{m}\frac{M_{\rm tot}}{\Sigma \pi p^2}T\label{eq:t_inc,dyn}\\
\tau_{\rm a, dyn} &=& \frac{(1-e^2)\sin I(\delta-\cos I)^{3/2}}{\sqrt{2}\bar{\kappa}|\cos I -\bar{\zeta}|}\frac{M_{\rm tot}}{m}\frac{M_{\rm tot}}{\Sigma \pi p^2} T\\
\tau_{\rm e, dyn} &=&  \frac{2e^2}{(1-e^2)}\left|\frac{1}{\tau_{\rm a, dyn}}-\frac{1}{\tau_{\rm p, dyn}}\right|^{-1}
\end{eqnarray}
}
\begin{figure}
	\includegraphics[width=\columnwidth]{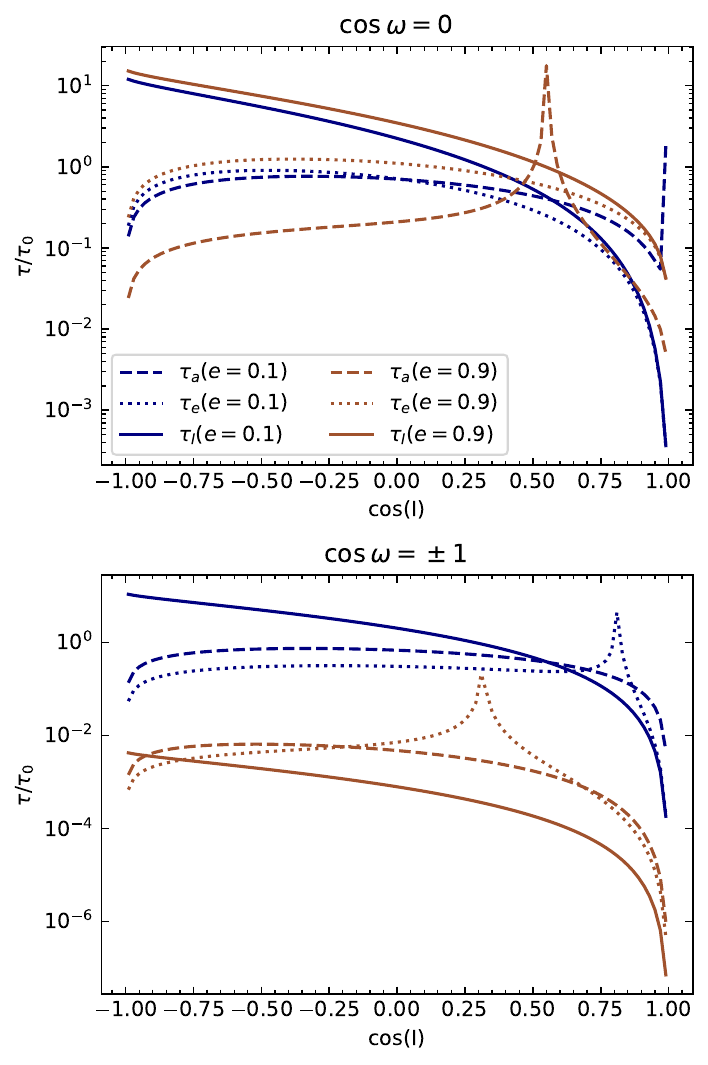}\\
    \caption{Timescales of semi-major axis, eccentricity and inclination evolution in a unit of $\tau_0=T\frac{M^2_{\rm tot}}{m\Sigma\pi p^2}$ {for BHs}.}
    \label{fig:tau-dyn}
\end{figure}
Figure~\ref{fig:tau-dyn} shows the timescale of semi-major axis, eccentricity, and inclination evolution in a unit of $\tau_0=T\frac{M_{\rm tot}}{m}\frac{M_{\rm tot}}{\Sigma\pi p^2}$. The upper panel shows the case for $\cos\omega=0$. Similar to aerodynamical drag, the timescale for the semi-major axis is much shorter than the timescales of inclination damping at high inclination and becomes longer as the inclination decreases. The timescales for $e$ and $I$ are insensitive to eccentricity. The bottom panel shows the case for $\cos\omega=\pm1$. For low eccentricity orbits that start from high inclination as indicated by blue lines, the eccentricity excitation timescale is much shorter than the semi-major axis/inclination damping. As the eccentricity becomes higher, the timescale for inclination damping becomes shorter, then the orbit inclination quickly damps, until the orbit becomes prograde, once the inclination is below the critical inclination given by Equation~\ref{eq:ie}, the eccentricity starts to decay. As eccentricity decreases, the timescale for inclination damping becomes longer. The semi-major axis damping takes over the inclination damping. As the inclination decreases further, the semi-major axis decay stops. Then the inclination and eccentricity timescales become shorter again at a very low inclination. A fully circularized orbit will be obtained at the end of the capture process.

\subsubsection{Integral of motion}
For Equation~\ref{eq:dLdI2}, similar to Section~\ref{sec:aero}, we can obtain the integral of motion,
\begin{eqnarray}
L\cos^2(I/2)\cot^{\zeta-1}(I/2)=\rm const\label{eq:LI2}
\end{eqnarray}
assumed that $\zeta$ is nearly constant within one order of magnitude as shown in Figure~\ref{fig:zeta}. The difference is that for the same $\cos\omega$ and $e$, the value of $\zeta$ is slightly smaller than $\gamma$.

The general picture for BHs is very similar to the stars as shown in Figure~\ref{fig:carton-star}. However, there's no parameter space for TDE and the eccentricity excitation timescale is much shorter than the stars. Therefore, the retrograde BHs will undergo significant eccentricity excitation. Meanwhile, the inclination-damping timescale for the BH is much more sensitive to the eccentricity, unlike the star the inclination-damping timescale is nearly constant. High eccentricity orbits decrease inclination much faster than circular orbits. Therefore, all the BHs will be eventually captured by the disk in prograde orbits. Since all the orbits will enter the fast eccentricity damping regime, they will eventually be fully circularized.

\begin{figure*}
    \includegraphics[width=2\columnwidth]{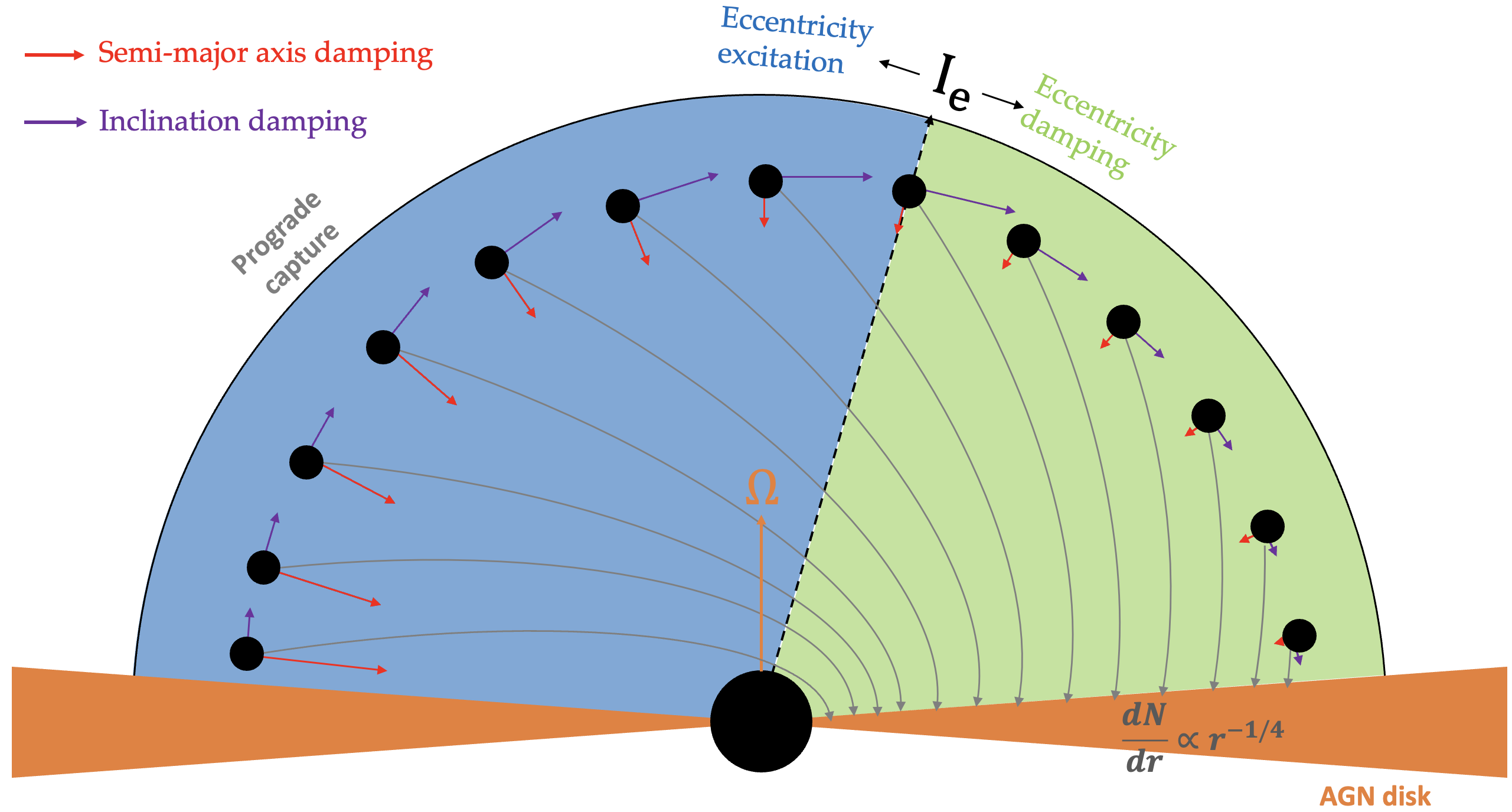}\\
    \caption{Carton shows the capture of BHs with different initial inclinations at a given fixed semi-major axis. For orbiters with $I>I_{\rm e}$, the orbital eccentricity will be excited. However, as the inclination decreases, the orbiter will enter a regime where eccentricity damping is fast. The trajectories and critical angles are not to scale, especially the semi-major axis that shrinks fast than the inclination damping and eccentricity evolution.}
    \label{fig:carton-bh}
\end{figure*}
Figure~\ref{fig:carton-bh} illustrates the BH capture process. For initially retrograde BHs with an inclination higher than $I_e$, their eccentricity becomes excited as the inclination dampens to $I_e$. Once their orbital inclination is below $I_e$, their eccentricity starts to decrease. 
\section{N-body examples}
To validate the calculations presented in Section~\ref{sec:calc}, we conducted N-body simulations using the software tool {\tt SpaceHub}\citep{Wang2021code}. Two sets of simulations were performed to examine the capture process involving a black hole (BH) and a main sequence star, where gas dynamical friction and aerodynamic drag dominate, respectively. In both cases, the mass of the supermassive black hole (SMBH) was set to $10^8M_\odot$, and the mass of the star or BH was set to $30M_\odot$. The initial semi-major axis was fixed at 0.1 pc for stars and 0.01 pc for BHs, and initial eccentricity was set to thermal average value $\sim 0.67$\footnote{We ran other initial eccentricities as well, but found that the trajectories are very similar. The evolution of the stars can be well described by equations in Section~\ref{sec:calc}.}. The initial inclinations were evenly distributed between 5 and 175 degrees, with an interval of 17 degrees. Each case was simulated with two different values of $\cos\omega$: 0 and {$\pm$}1.

For the star, the force was implemented according to Equation~\ref{eq:f_aero}, while for the BH, the drag force was implemented as described in Equation~\ref{eq:f_dyn}. The Coulomb logarithm $\Lambda$ was kept constant at 3 for gas dynamical friction.

\subsection{Disk models}\label{sec:disk}
We adopted two different disk models for the capture process, a thin $\alpha$-disk model that can be well described by simple equations and a thick Sirko-Goodman (SG) disk model that has larger surface density around the active capture radius.
\subsubsection{$\alpha$-disk}

The $\alpha$ disk model can be characterized by three parameters: the accretion rate efficiency ($\lambda_{\rm d}$), the Toomre parameter ($Q_{\rm d}$), and the viscosity parameter ($\alpha_{\rm d}$). The accretion rate can be approximated as follows:
\begin{eqnarray}
\dot{M}&=&\lambda_{\rm d} \dot{M}_{\rm Edd} = 2.2\lambda_{\rm d}\left(\frac{M_{\rm SMHB}}{10^8M_\odot}\right) M_\odot/{\rm yr}\\
\dot{M}_{\rm Edd} &=& \frac{L_{\rm Edd}}{\eta_{\rm d} c^2}
\end{eqnarray}
where, $\dot{M}$ represents the accretion rate, $\dot{M}_{\rm Edd}$ denotes the Eddington accretion rate, $M_{\rm SMHB}$ represents the mass of the SMBH, $L_{\rm Edd}$ is the Eddington luminosity, and $\eta_{\rm d}$ is a constant with a value of 0.1.

The disk surface density $\Sigma$ can be described by,
\begin{eqnarray}
\Sigma &=& \frac{\dot{M}}{2\pi r v_r}\\
v_r &=& \alpha_d h^2 \sqrt{\frac{GM_{\rm SMBH}}{r}}\\
h &=& H / r = \left(\frac{Q_{\rm d}\dot{M}}{2\alpha_{\rm d}M_{\rm SMBH}\Omega_{\rm d}}\right)^{1/3}
\end{eqnarray}
where, $r$ represents the distance from the SMBH, $H$ is the scale height, and $\Omega_{\rm d}$ denotes the orbital frequency given by $\sqrt{GM_{\rm SMBH}/r^3}$. The parameters $\lambda_{\rm d}$, $\alpha_{\rm d}$, and $Q_{\rm d}$ are assumed to be constant values set to 1.
\subsubsection{Sirko-Goodman (SG) disk}
The solution of the SG model\citep{Sirko2003} can only be solved numerically. There are no simple equations that can be used directly to describe the SG disk profile.

\begin{figure}
    \includegraphics[width=\columnwidth]{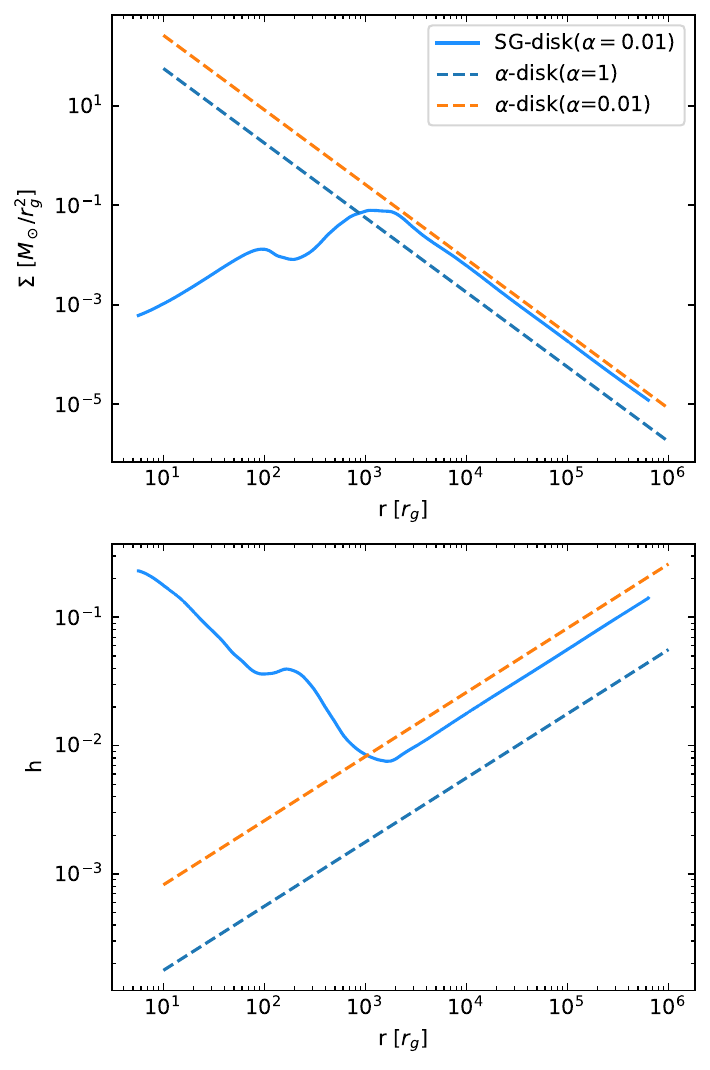}
    \caption{Surface density and specific scale height of $\alpha$-disk and SG disk.}
    \label{fig:disk}
\end{figure}
{Figure~\ref{fig:disk} shows the surface density and specific scale height of the $\alpha$-disk and SG disk around a $10^8$ $M_\odot$ SMBH. the surface density and specific scale height of the SG model with $r>1.2\times10^3r_g$ is roughly 3 times of the surface density of an $\alpha$-disk model with $\alpha=1$. In the inner region, the surface density of the $\alpha$-disk is much higher than the SG disk { since the inner disk in the SG model is gravitationally stable}. The surface density and specific scale height of the SG disk in the outer disk regime are similar to an $\alpha$-disk with $\alpha=0.01$.} 

\subsection{Stellar trajectories}
\begin{figure*}
    \includegraphics[width=2\columnwidth]{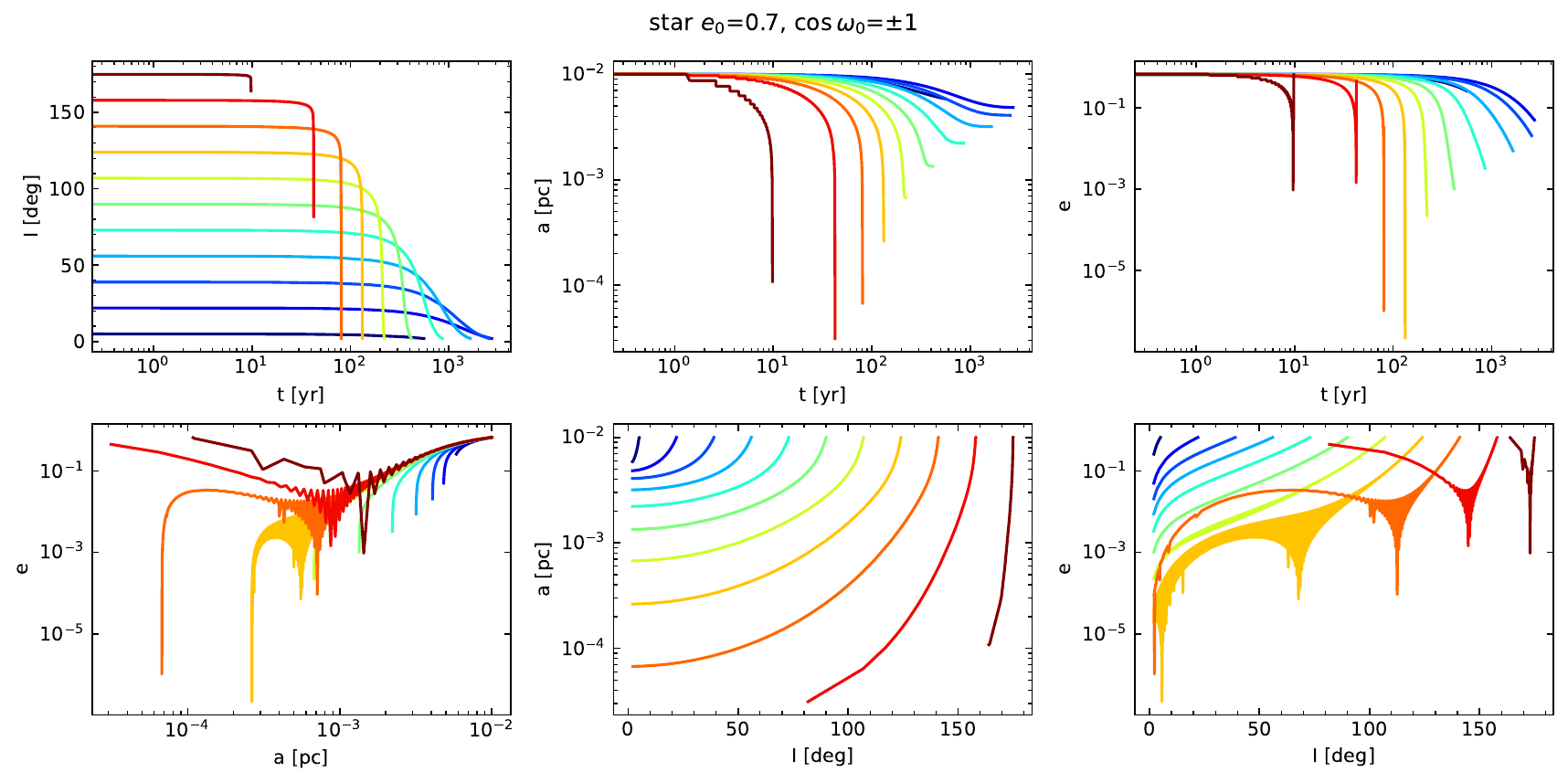}
    \caption{Trajectories of stars for different initial inclinations. Upper panels: inclination, semi-major axis and eccentricity as a function of time. Bottom panels: $e$ as a function of $a$, $a$ as a function of $I$ and $e$ as a function of $I$. Different colors indicate different initial inclinations that can be seen in the upper left panel. }
    \label{fig:star_traj}
\end{figure*}
Figure~\ref{fig:star_traj} depicts the trajectories of a 30 solar mass star orbiting an SMBH with a fixed initial thermal average eccentricity of 2/3 $\sim$ 0.67 and various initial inclinations. All orbits are assumed to have $\cos\omega=\pm1$ initially.

The upper middle panel of the figure demonstrates that for orbits with high inclinations, the semi-major axis experiences faster decay, as described by Equation~\ref{eq:t_a_aero} and Figure~\ref{fig:tau-aero}. The inclination damping timescale is much longer than the semi-major axis damping timescale, as observed in the bottom middle panel. Consequently, stellar orbits shrink rapidly while inclination damping occurs at a slower rate.

As the inclination decreases, the semi-major axis damping timescale increases significantly until it surpasses the inclination damping timescale. This results in the stalling of the semi-major axis decay and a pronounced decrease in inclination, as shown in the bottom panel of Figure~\ref{fig:tau-aero}.

The upper right panel displays the eccentricity evolution. In agreement with Equation~\ref{eq:dedt} and the bottom panel of Figure~\ref{fig:tau-aero}, both prograde and  retrograde orbits experience eccentricity damping. As illustrated in the bottom right panel, orbits with very high inclinations exhibit a significantly shorter timescale for eccentricity excitation compared to the inclination-damping timescale. However, as the semi-major axis rapidly decreases, the aerodynamic drag becomes very efficient due to the high disk surface density. The quasi-static assumption that the orbital parameter changes per period are small, which we adopted in the derivation, is no longer true. Thus, for retrograde orbits with high inclinations, we observe some unpredicted eccentricity excitations.

The upper left panel presents the inclination evolution in the presence of semi-major axis damping. As the semi-major axis decreases, the disk's surface density increases substantially, leading to a much shorter inclination-damping timescale. Since high-inclination orbits have shorter semi-major axis damping timescales, orbits with high inclinations are captured more rapidly by the disk.

The trajectories with other initial eccentricities and $\cos\omega$ are very similar to Figure~\ref{fig:star_traj} and can be well described by timescales in Section~\ref{sec:aero}, Figure~\ref{fig:tau-aero} and Figure~\ref{fig:carton-star}.

\subsection{BH trajectories}
\begin{figure*}
    \includegraphics[width=2\columnwidth]{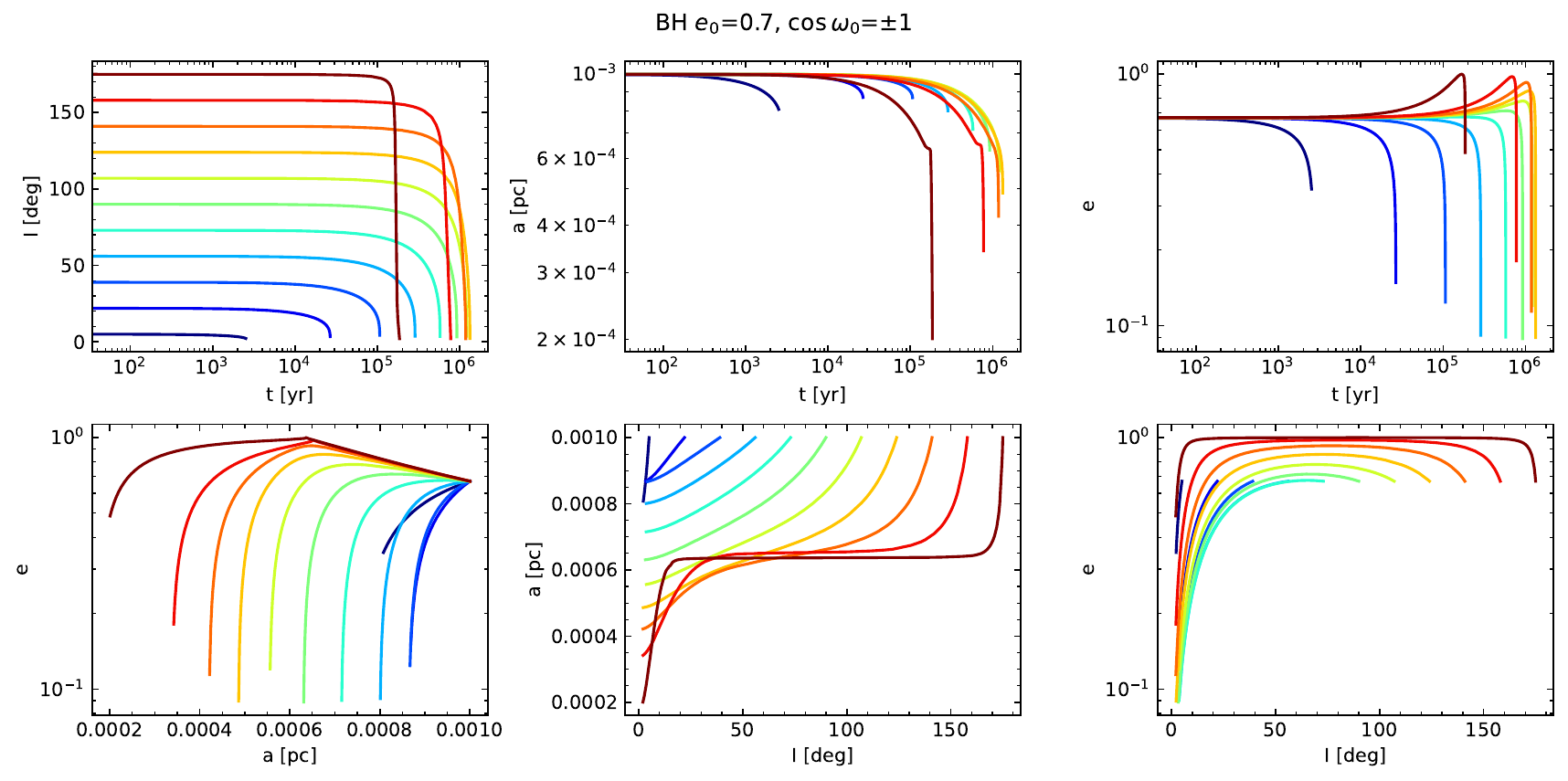}
    \caption{ Similar to Figure~\ref{fig:star_traj}, but for 30 $M_\odot$ BHs.}
    \label{fig:BH_traj}
\end{figure*}

Figure~\ref{fig:BH_traj} shows the trajectories of 30 $M_\odot$ BHs starting with different inclinations. For large initial inclinations, because the eccentricity timescale is much shorter than the semi-major axis and inclination damping timescale as indicated by Figure~\ref{fig:tau-dyn}, and the $\dot{e}$ is positive, therefore, the eccentricity of highly inclined orbits grows fast. As the eccentricity becomes high, the inclination damping timescale of those eccentric orbits becomes very short, leading to fast inclination decreases. Once the orbital inclination is below the critical angle $I_{\rm e}$, the eccentricity starts to decrease. For retrograde BHs, they all will undergo thus an eccentricity excitation and damping process during the interaction with the disk. Unlike the stars, whose inclination damping timescale is insensitive to the eccentricity, the inclination damping timescale for BH is very sensitive to the eccentricity. Therefore for retrograde BHs, the capture process can be divided into three stages: fast eccentricity growth, quick inclination damping and slow eccentricity damping.  The semi-major axis keeps shrinking along the three stages. Since the eccentricity damping may not be as efficient as inclination damping, there might be some eccentricity residual on captured BHs. 

For initially prograde BHs, the picture is very similar to the stars, except the inclination and eccentricity evolution timescale is relatively shorter than the semi-major axis damping timescale. Due to this reason and the residual eccentricity in the integral of motion $a(1-e^2)\cos^4(I/2)$, the semi-major axis of captured BHs may not shrink as much as stars.

\subsection{Test the integral of motion}
To verify the integral of motion we obtained in Section~\ref{sec:c of motion}, we set up Monte Carlo (MC) simulations with different initial orbital conditions and tracked the orbital parameters of those orbits during the disk capture process. The mass of the black hole (BH) and star is kept constant at 50$M_\odot$. The semi-major axes are drawn from a distribution such that the density profile of the nuclear star cluster (NSC) follows $\propto r^{-1.5}$ from 1$r_g$ to 0.1 pc. The eccentricities are drawn from a thermal distribution with a probability density function $p(e)=2e$, and the inclinations are drawn from a distribution where $\cos I$ is uniformly distributed between -1 and 1. We use the $\alpha$-disk model for the MC simulations. The MC simulations are terminated once the inclination of the orbiter is below the specific scale height of the disk.
\begin{figure}
    \includegraphics[width=\columnwidth]{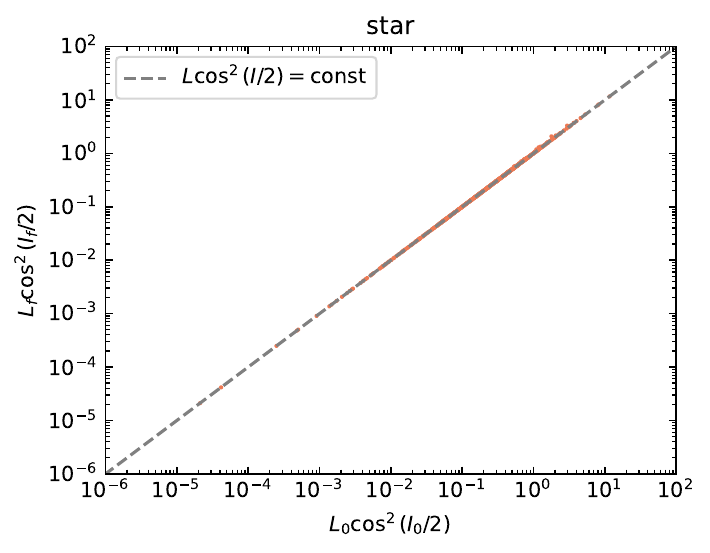}\\
    \includegraphics[width=\columnwidth]{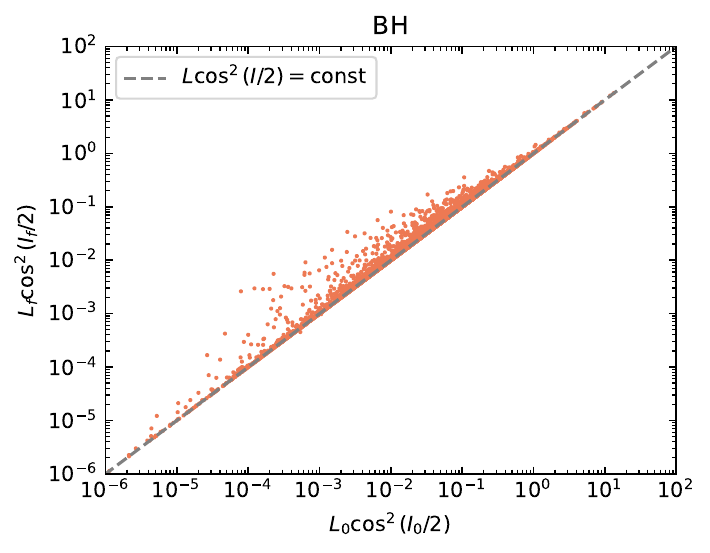}
    \caption{The initial value verses the final value of the integral of motion for the numerical simulations. The upper panel shows the cases for star and the bottom panel shows the cases for BH.}
    \label{fig:verify}
\end{figure}
Figure~\ref{fig:verify} shows the integral of motion at the beginning and end of the MC simulations. The upper panel shows the results for stars, while the bottom panel displays the results for black holes (BHs). As depicted in Figures~\ref{fig:gamma} and \ref{fig:zeta}, $\zeta$ has a much larger range than $\gamma$. Therefore, according to Equation~\ref{eq:LI} and \ref{eq:LI2}, it is expected that $L\cos^2(I/2)$ will deviate more for the BH. One should note that this integral of motion only holds true for the capture process. Once the objects become fully embedded within the disk, this integral of motion is no longer a constant value.

\section{Populations of captured objects}
Due to the significant observational challenges, the profile, compact object fraction, and binary fraction in the nuclear star cluster (NSC) are poorly constrained. Additionally, the capture process in AGN disks can be intrinsically complex with numerous uncertainties due to the variety of AGN disk models. Consequently, establishing the population of captured objects in AGN disks can be a challenging task that highly depends on the adopted NSC and disk models.

However, in this section, we will demonstrate that certain properties of the captured objects remain insensitive to the NSC and disk models. This suggests that it is possible to establish a robust population of captured objects in AGN disks.

As discussed earlier, due to the short timescale of eccentricity damping, both stars and black holes (BHs) will be captured in nearly circular orbits. Using the integral of motion, the final captured semi-major axis can be well described as $a_{\rm f} = a_0(1-e_0^2)\cos^4(I_0/2)$. Furthermore, since the inclination damping timescale, $\tau_I$, is insensitive to the inclination, eccentricity, and argument of periapsis of the orbit, we can obtain the distribution of captured objects in AGN disks at any given time for a given disk model and NSC profile.

\subsection{Density power-law index for captured objects}
We employ the same disk models as in Section~\ref{sec:disk}, namely an $\alpha$-disk with reasonable disk parameters $Q_{\rm d}=1$, $\lambda_{\rm d}=1$, and $\alpha_{\rm d}=0.1$, as well as a SG disk model. For the NSC model, we consider two different masses: 1 $M_\odot$ and 50 $M_\odot$ for stars, and 5 $M_\odot$ and 50 $M_\odot$ for black holes (BHs). The number density/density profile of the NSC follows a power-law distribution, where the number density is given by \citep{Merritt2013}:

\begin{eqnarray}
n = \frac{3-\gamma_{\rm NSC}}{2\pi}\frac{M_{\rm SMBH}}{m}r_{\rm m}^{-3}\left(\frac{r}{r_{\rm m}}\right)^{-\gamma_{\rm NSC}}\label{eq:n-nsc}
\end{eqnarray}

Here, $m$ represents the mass of the star/BH in the NSC, and $r_{\rm m}$ corresponds to the gravitational influence radius of the supermassive black hole (SMBH). We adopt three power-law index $\gamma_{\rm NSC}$, 1, 1.5 and 2. Within this radius, the total enclosed masss is 2$M_{\rm SMBH}$ and the dynamics of orbits is predominantly influenced by the SMBH's gravity, allowing us to neglect the effects of the dark matter halo and other stars/BHs. The value of $r_{\rm m}$ can be determined using the following equation:

\begin{eqnarray}
r_{\rm m} = \frac{GM_{\rm SMBH}}{\sigma_{\rm NSC}^2}
\end{eqnarray}

where $\sigma_{\rm NSC}$ represents the velocity dispersion of the NSC, given by \citep{kormendy2013}:

\begin{eqnarray}
\sigma_{\rm NSC} = 2.3 \rm km/s \left(\frac{M_{\rm SMBH}}{M_\odot}\right)^{{1}/{4.38}}
\end{eqnarray}

\begin{figure*}
\includegraphics[width=2\columnwidth]{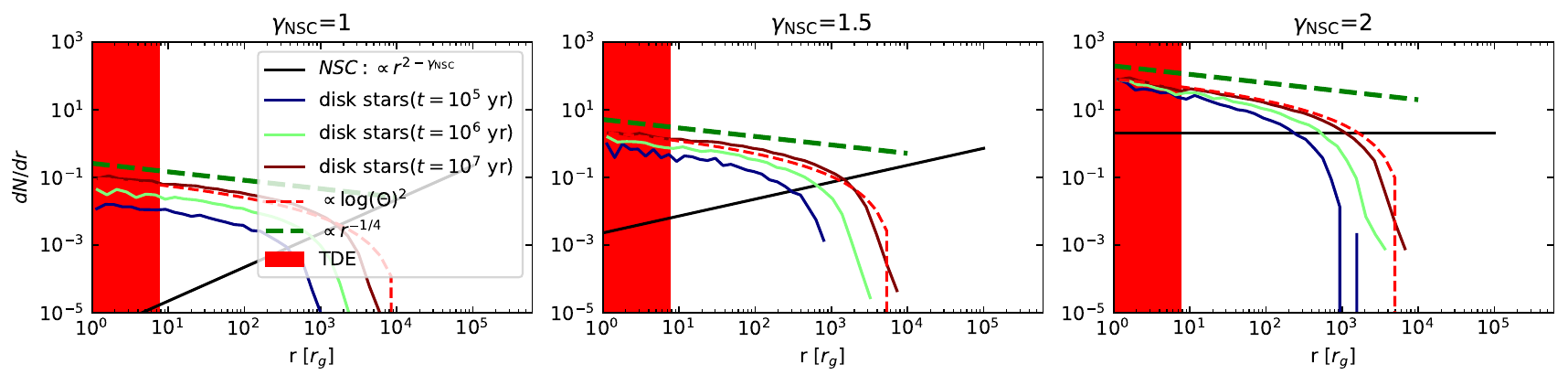}\\
    \includegraphics[width=2\columnwidth]{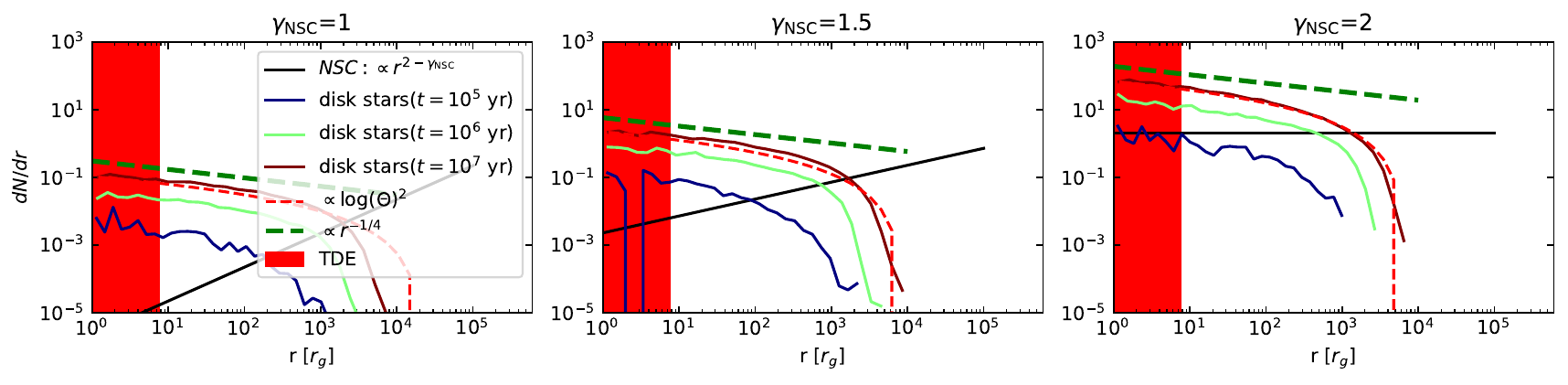}
    \caption{$\dd N/\dd r$ as a function of $r$ for the initial NSC consists of 50 $M_\odot$ stars and captured stars in the disk at different times. The upper panels show the results for $\alpha$-disk (with $\alpha=0.1$) capture and the bottom panels show the results for SG disk capture. Different columns indicate different initial NSC density profiles. The NSCs are set to be isotropic with number density profile dn/dr $\propto r^{-\gamma_{\rm NSC}}$. Regardless of the NSC density profile and disk model, the density profile of the captured star follows $\dd N/\dd r$ $\propto r^{-1/4}$.}
    \label{fig:dNdt-star}
\end{figure*}
To obtain the population of captured objects in AGN disks at any given time $t$, we assume that objects in the NSC will be captured by the disk after, the inclination damping timescale ($t=\tau_{\rm I}$). Once objects are captured by the disk, we populate the captured objects at the disk mid-plane. Based on the integral of motion during the capture process, we can calculate the final semi-major axis from $a_{\rm f} = a_0(1-e_0^2)\cos^4(I_0/2)$. Figure~\ref{fig:dNdt-star} depicts the 1-D number density profiles of captured 50 $M_\odot$ stars at different times for various initial NSC profiles and disk models. The upper panels present the number density profiles of captured disk stars for the $\alpha$-disk, while the lower panels display the profiles for the SG disk. Each column corresponds to different NSC profiles. Surprisingly, the number density profiles of the captured disk stars exhibit remarkable consistency across different disk models and NSC profiles.

\begin{figure*}
    \includegraphics[width=2\columnwidth]{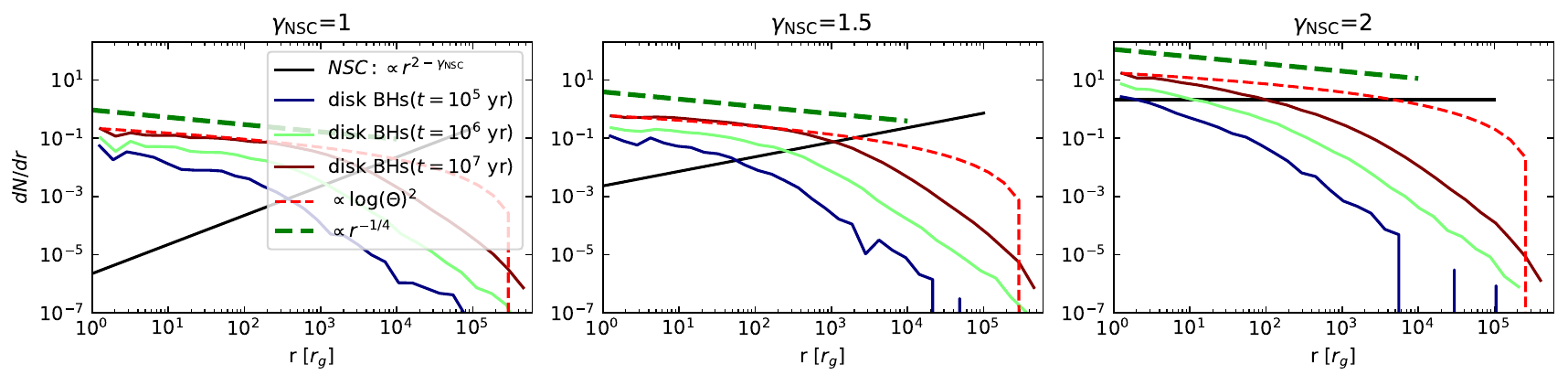}\\
    \includegraphics[width=2\columnwidth]{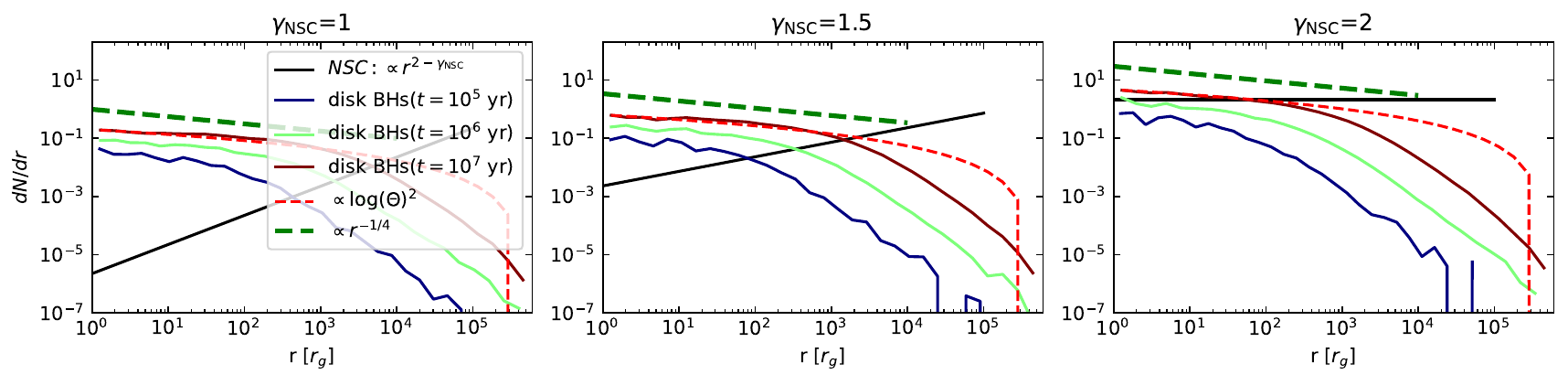}
    \caption{Similar to Figure~\ref{fig:dNdt-star}, but for the NSC consists of 50 $M_\odot$ BH.}
    \label{fig:dNdt-BH}
\end{figure*}

The captured disk stars follow a density profile proportional to $r^{-1/4}$ from $10^5$ years, which represents the typical lifetime of short-lived AGN disks, to $10^8$ years for long-lived AGN disks. The capture process causes an accumulation of stars in the inner region of the disk. The profile of the disk stars is not sensitive to the NSC radial profiles because the final captured semi-major axis is more sensitive to the inclination and eccentricity distribution of the orbits within the NSC, owing to the angular dependence in the integral of motion $L\cos^2(I/2)$. As we assume the NSC to be nearly isotropic and the eccentricity is isothermal, the captured disk star profiles exhibit a high level of consistency. Indeed, in the asymptotic regime, if we assume all the orbits can be captured by the disk, the final semi-major axis of the capture objects is $a_{\rm f} = a_0(1-e_0^2)(\frac{1+\cos I}{2})^2$. Since both $1-e_0^2$ and $(1+\cos I)/2$ are uniformly distributed between [0,1], therefore the probability function of $\Theta = (1-e_0^2)(\frac{1+\cos I}{2})^2$ is
\begin{eqnarray}
p(\Theta)=\frac{1}{2}\ln^2(\Theta), \Theta\in [0,1]
\end{eqnarray}
Then the probability function for $a_{\rm f}$ can be obtained from
\begin{eqnarray}
p(a_{\rm f})&\propto& \int_0^{r_{\rm max}} p(a_0)\ln^2(\Theta = a_{\rm f}/a_0) \dd a_0\nonumber\\
  &\propto& \frac{r_{\rm max}^{3-\gamma_{\rm NSC}}}{3-\gamma_{\rm NSC}}\ln^2(a_{\rm f}/a_0), a_{\rm f}\in [0,a_0]
\end{eqnarray}
where $p(a_0)$ is the power-law probability function of the initial semi-major axis distribution in NSC and $r_{\rm max}=r_{\rm m}$ is the outer boundary of the NSC.
The dashed lines in Figure~\ref{fig:dNdt-star} and \ref{fig:dNdt-BH} indicate that this probability function that follows nearly $\propto r^{-1/4}$ (asymptotic to $\ln^2(r)$ in the inner region) is a very good approximation for the final semi-major axis distribution. The captured star/BH number density profile thus depends on the angular distribution of the star/BH in NSC rather than the radial distribution. An isotropic star/BH distribution in NSC gives $\dd N/\dd r$ $\propto \sim r^{-1/4}$.

Figure~\ref{fig:dNdt-BH} displays the disk star profile for the 50 $M_\odot$ black holes (BHs). Since the integral of motion for BHs deviates slightly from $L\cos^2(I/2)$, the captured BH profile follows a constant distribution. In contrast to star capture, the density profiles for BHs {extend to further out of the AGN disks}. This indicates that BHs can be captured in the outer region of the disk up to $10^6r_g$.

\begin{figure}
    \includegraphics[width=\columnwidth]{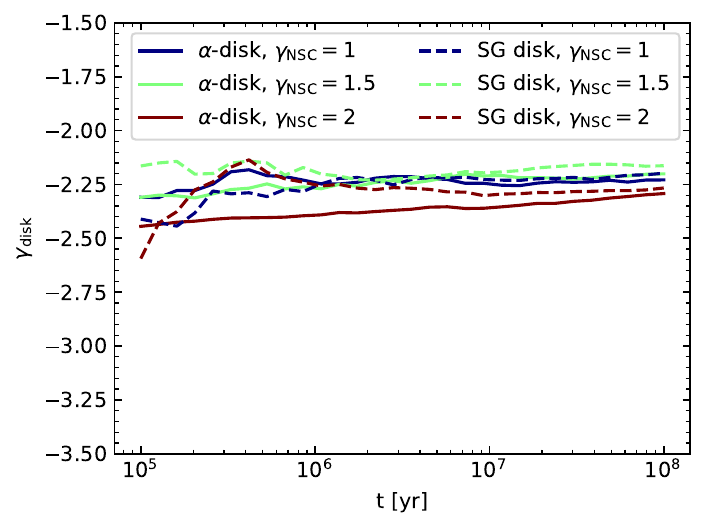}
    \includegraphics[width=\columnwidth]{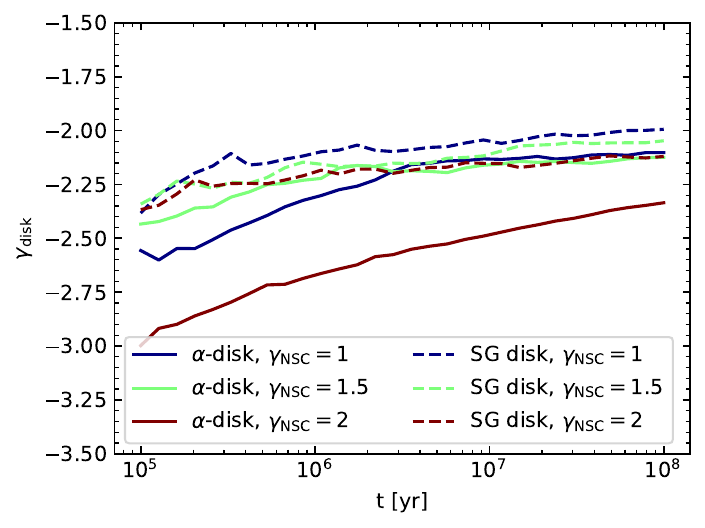}
    \caption{Density profile power-law index of captured stars/BHs as a function of time for different disk models and initial NSC profile. Both the upper panel (stars) and bottom panel (BG) show a converged power-law index.}
    \label{fig:index}
\end{figure}
Figure~\ref{fig:index} illustrates the power-law index of the captured disk stars/BHs as a function of time for different disk models and NSC density profiles. In the upper panel, the power-law index of the 3D number density profile for captured disk stars exhibits remarkable consistency and converges to approximately -2.25, indicating $dN/dt \propto r^{-1/4}$.

\subsection{Mass filling function of captured objects }\label{sec:m-fill}
{Besides the initially embedded objects}, the total captured mass can be obtained by
\begin{eqnarray}
\mathcal{M}(t) =f_*\int_{\tau_{\rm I}<t}2\pi p(\cos I)p(e)p(a)a^2\dd a \dd e \dd\cos I
\end{eqnarray}
where $f_*$ is the mass fraction of the star in NSC, $p(e)=2e$ is the probability function for eccentricity distribution, $p(\cos I)=1/2$ is the angular distribution function and $p(a)$ is the probability function for density distribution in NSC that can be obtained by multiplying $m$ to Equation~\ref{eq:n-nsc}.

For captured stars, the total captured mass by the AGN disk can be approximated as
\begin{eqnarray}
\mathcal{M}_*(t) &\sim& 2M_{\rm SMBH}f_* \bigg(\frac{\Sigma_m}{\Sigma_*}\frac{t}{T_m}\bigg)^{1-\gamma_{\rm NSC}/3}
\end{eqnarray}
where $T_m = 2\pi\sqrt{\frac{r_m^3}{GM_{\rm SMBH}}}$ and $\Sigma_m$ are the orbital period and surface density at the gravitational influence radius $r_m$, respectively\footnote{We assume the surface density scale with $\propto r^{-3/2}$, which is true for self-gravitating disk models with constant accretion rate.}. The captured stars distribute between $R_{\rm TDE}$ and $r_{\rm max,*}$ with a power-law distribution $\dd N/\dd t\propto r^{-1/4}$, where
\begin{eqnarray}
r_{\rm max,*} \sim r_m\left(\frac{\Sigma_m}{\Sigma_*}\frac{t}{T_m}\right)^{1/3}
\end{eqnarray}
$\mathcal{M}_*=2M_{\rm SMBH}f_*$ indicates all stars within $r_m$ (with the total mass of 2$M_{\rm SMBH}$) are captured by the disk. The corresponding time $t = T_m\Sigma_*/\Sigma_m$ is exactly the capture timescale for a circular orbit at $r_m$, the final captured orbit of the NSC.

For captured BHs, the total captured mass can also be obtained as
\begin{eqnarray}
\mathcal{M}_{\rm BH}(t) &\sim&\bigg[ \left(\frac{m}{M_{\rm SMBH}}\frac{\Sigma_m\pi r_m^2}{M_{\rm SMBH}}\frac{t}{T_m}\right)^{3-\gamma_{\rm NSC}} \int_{-1}^{\mu}\frac{\dd\cos I}{(I\sin(I/2))^{3-\gamma_{\rm NSC}}} \nonumber\\
&+& \int_\mu^{1}\dd\cos I\bigg]2M_{\rm SMBH}f_{\rm BH}\nonumber \\
&\sim&2M_{\rm SMBH}f_{\rm BH}\frac{m}{M_{\rm SMBH}}\frac{\Sigma_m\pi r_m^2}{M_{\rm SMBH}}\frac{t}{T_m}
\end{eqnarray}
where $\mu\sim 1-\frac{m}{M_{\rm SMBH}}\frac{\Sigma_m\pi r_m^2}{M_{\rm SMBH}}\frac{t}{T_m}$. The captured BHs are populated between $r_g$ and $r_m$ with a roughly power-law distribution $dN/dt\propto r^{-1/4}$.

\begin{figure}
    \includegraphics[width=\columnwidth]{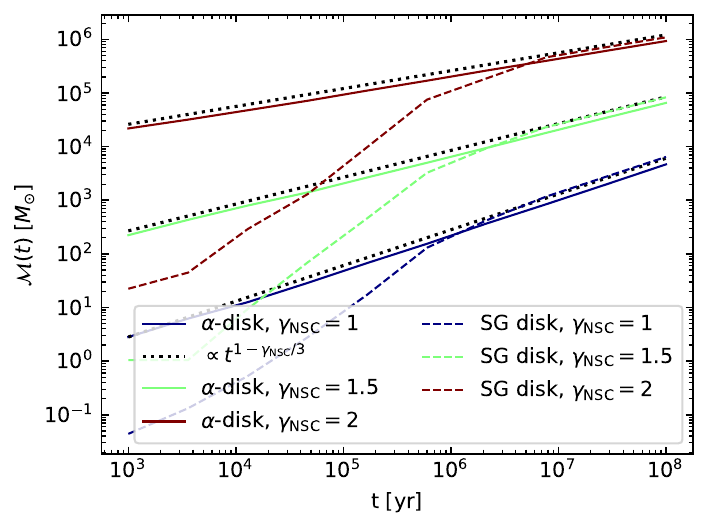}
    \includegraphics[width=\columnwidth]{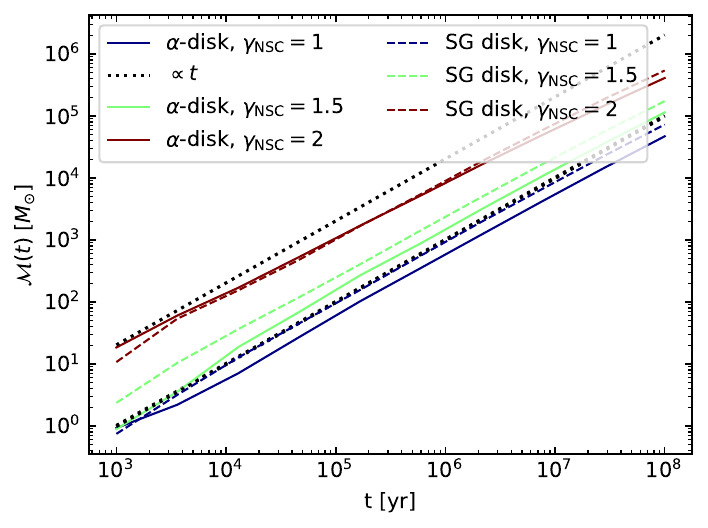}
    \caption{Total mass of captured objects as a function of time for different disk models and NSC density profile. The upper panel shows the results for the captured star and the bottom panel shows the results for captured BHs}
    \label{fig:M}
\end{figure}
Figure~\ref{fig:M} shows the captured total mass as a function of time for different disk models and NSC density profiles from our MC simulations. The upper panel shows the cases for stars with $f_*=1$ and the bottom panel shows the cases for BHs with $f_{\rm BH}=1$. As indicated, the mass filling function for stars $\mathcal{M}_*\propto t^{1-\gamma_{\rm NSC}}$ and the mass filling function for BHs $\mathcal{M}_{\rm BH}\propto t$.

\subsection{Direct binary formation from capture}
The formation of binary black holes (BHs) in AGN disks relies on migration traps, where single BHs can be captured and accumulated. In the migration trap, the net torque resulting from Lindblad resonance is zero, allowing BHs to be trapped and accumulate in that region. However, migration traps strongly depend on the specific disk models. Furthermore, the concept of migration traps is based on the single migrator model, where the gravitational effects of other migrators can be ignored. In reality, in an N-body migration system, resonances from other migrators come into play and can significantly influence the migration of individual BHs.

Considering that the population of captured stars and BHs can be well described by a power-law distribution regardless of the NSC density profiles and disk models, we can examine how dynamically crowded the captured objects become once they are captured. We employ a quantity called $N_{\rm Hill}$, which represents the number of objects within the volume of a ring with a radius $r$ and a cross-section radius $R_{\rm Hill}=r(\frac{m}{M_{\rm SMBH}})^{1/3}$. This number indicates how many objects can enter the Hill's radius of a single orbiter over time. If $N_{\rm Hill}$ is larger than two, it suggests that a binary could potentially form through encounters aided by gas dissipation in AGN disks. Conversely, if $N_{\rm Hill}$ is too small, it indicates that the embedded objects are dynamically distant from each other.

\begin{figure*}
    \includegraphics[width=2\columnwidth]{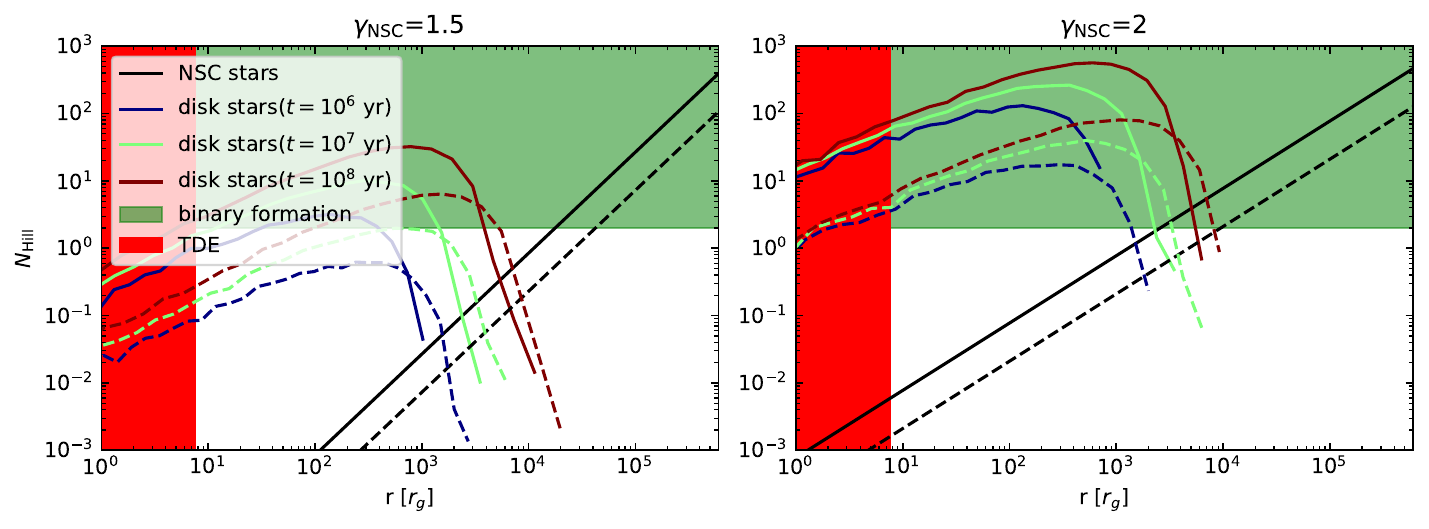}
    \includegraphics[width=2\columnwidth]{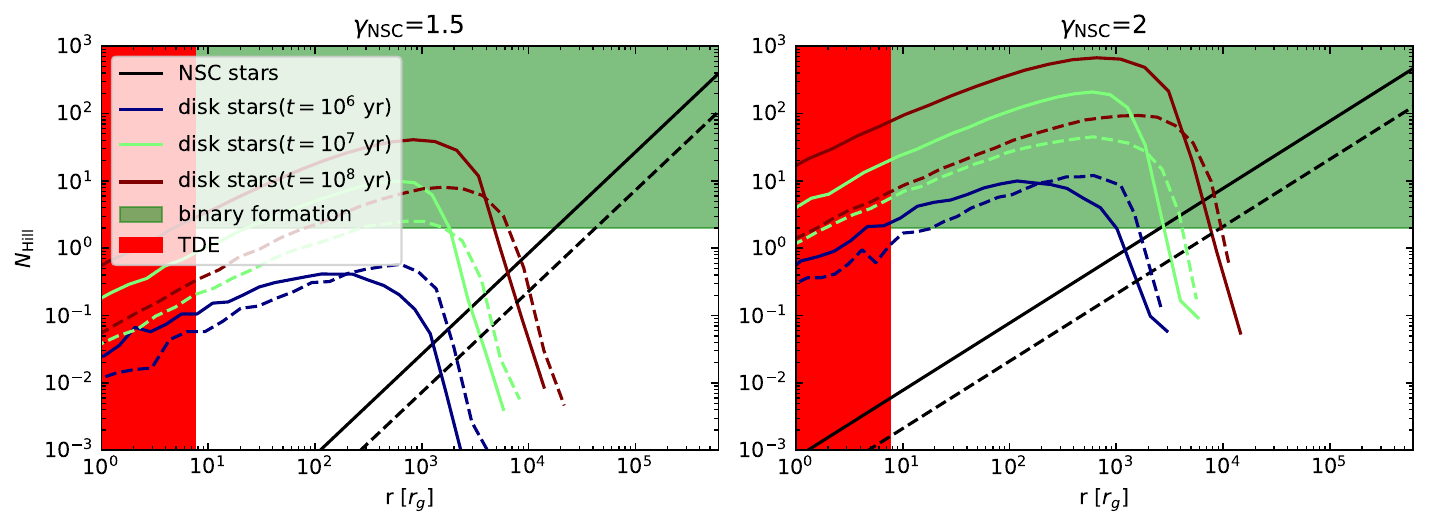}
    \caption{Number of objects within the Hill's ring for initial NSC and captured stars at different radii and at different times. A number of objects greater than two indicate possible binary formation. Upper panels: objects are captured by a $\alpha$-disk (with $\alpha=0.1$); Bottom panels: objects are captured by a SG disk; Left panels: the initial NSC density profile follows $\rho(r)\propto r^{-1.5}$; Right panels: the initial NSC density profiles follows $\rho(r)\propto r^{-2}$. The solid lines are for NSC consists of 1 $M_\odot$ stars and the dashed lines are for 50 $M_\odot$ stars.}
    \label{fig:Nhill-star}
\end{figure*}

Figure~\ref{fig:Nhill-star} depicts $N_{\rm Hill}$ as a function of $r$ for different disk models and NSC density profiles around a $10^8M_\odot$ SMBH. The upper panels show the results for an $\alpha$-disk with $\alpha = 0.1$, while the lower panels display the results for an SG disk with $\alpha = 0.01$. Solid lines represent $N_{\rm Hill}$ for an NSC consisting of 1 $M_\odot$ stars (with higher number density), while dashed lines represent 50 $M_\odot$ stars (with lower number density).

Initially, $N_{\rm Hill}$ in the NSC is generally smaller than two, except for the outer layer at a distance of approximately $10^4r_g$ from the SMBH. However, the capture process reduces the inclination of orbits at a given $r$ and leads to the accumulation of captured stars in the radial direction, resulting in a dynamically crowded population. In the case of an $\alpha$-like disk with a relatively high surface density in the inner disk region and a relatively low surface density in the outer disk region, along with a not too steep NSC profile ($\gamma_{\rm NSC}=1.5$), only NSCs with very high number density consisting of 1 $M_\odot$ stars can be accumulated by AGN disks. If the number density of the NSC is low, only a long-lived AGN disk with a lifetime of approximately $10^8$ years can capture stars with $N_{\rm Hill}>2$. Notably, $N_{\rm Hill}$ increases significantly as the NSC density profile becomes steeper and insensitive to the disk models.

\begin{figure*}
    \includegraphics[width=2\columnwidth]{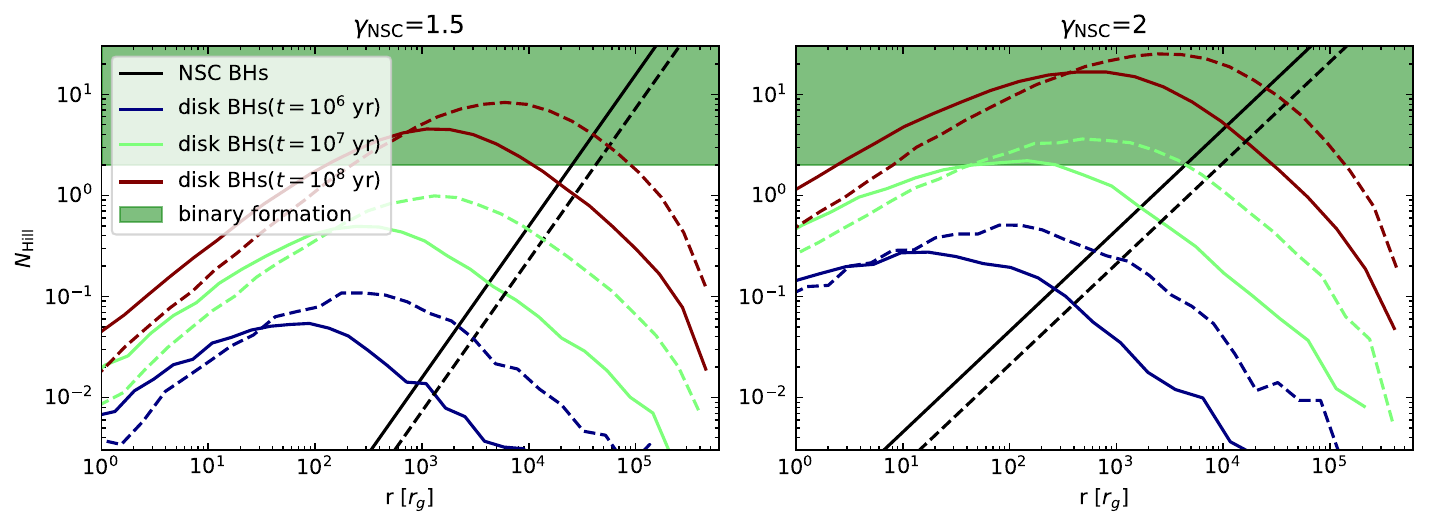}
    \includegraphics[width=2\columnwidth]{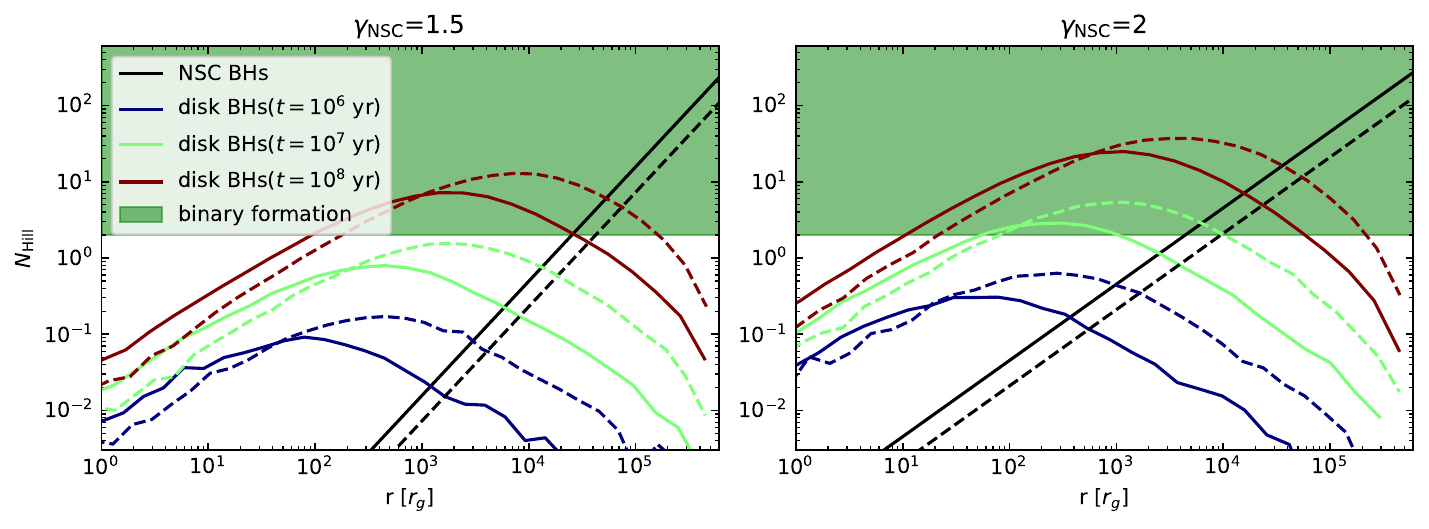}
    \caption{Similar to Figure~\ref{fig:Nhill-star}, but for BHs.}
    \label{fig:Nhill-BH}
\end{figure*}

Figure~\ref{fig:Nhill-BH} presents the results for BHs, similar to Figure~\ref{fig:Nhill-star}. Since the capture timescale (Equation~\ref{eq:t_inc,dyn}) is much longer for BHs compared to stars (Equation~\ref{eq:t_inc,aero}), on average, more time is required for the disk to accumulate BHs.  For NSC with mild power-law index $\gamma_{\rm NSC}=1.5$, only a long-lived AGN disk makes it possible to directly form binary BH from the capture process. Normal lifetime and short-lived AGN disk result in a dynamically distant BH population in AGN disks. 

For more concentrated NSCs, the direct BH binary formation is more promising. For a normal lifetime AGN disk, direct BH binary formation occurs around hundreds of $r_g$.

Although, in general, captured BHs are not dynamically crowded with $N_{\rm Hill}>2$. However, the captured stars are over-dense upon being captured by the AGN disk. There will be frequent binary star formations in the AGN disk around hundreds of $r_g$. Those binaries could eventually evolve into binary BHs through stellar evolution. The stellar evolution in the AGN disk is far more complicated that involves non-isotropic accretion, and internal stellar structure, which is beyond the scope of this paper {  \citep{Cantiello2021, Jermyn2022, alidib2023,
huang2023}}. However, we believe, in this over-dynamical-dense region, there will be binary BH remaining throughout the stellar evolution. Those binary BH could be the source of merging BHs in AGN disks and bypass the traping phase with lots of uncertainties from N-body migration. 

\section{Conclusions}
We use analytical and numerical methods to investigate the AGN disk-nuclei star cluster interaction process that is crucial for stellar evolution and BH mergers in AGN disks. The conclusions we obtained can be summarised as follows,
\begin{itemize}
    \item For disk-star/BH interaction, we derived the timescale expressions for semi-major axis, eccentricity and inclination evolution for aerodynamic drag and gas dynamical friction. 

    \item Based on the derived timescales, in general, for star capture, the semi-major axis damping timescale is much shorter than the inclination damping timescale. The inclination damping timescale is also much shorter than the eccentricity evolution timescale. Therefore, for star capture, the capture process can be roughly divided into three stages: semi-major axis shrinking, eccentricity evolution, and inclination damping. These timescales are insensitive to initial eccentricities, but the semi-major axis and eccentricity timescales are very sensitive to inclinations. High inclination orbits have much shorter timescales for semi-major axis/eccentricity evolution compared to low inclination orbits.

     \item For BH capture, the eccentricity evolution and inclination damping timescales are sensitive to initial eccentricities, inclinations, and the orbital argument of periapsis with respect to the disk. There is no general conclusion as to which process is faster among the three. High inclination orbits can shrink their semi-major axis faster than low inclination orbits. Similarly, high eccentricity orbits can decrease their inclination faster than low eccentricity orbits.

     \item The eccentricity evolution is sensitive to the orbit argument of periapsis $\omega$ with respect to the disk. However, for most values of the argument of periapsis, high inclination (retrograde) orbits will undergo eccentricity excitation, while low inclination (prograde) orbits will undergo eccentricity damping. The critical angle $I_e$ that distinguishes the two regimes is derived. For orbits with an inclination higher than $I_e$, eccentricity excitation followed by eccentricity damping occurs as the inclination decreases to zero.

    \item The integrals of motion $L\cos^2(I/2)\cot^{\gamma-1}(I/2)$ and $L\cos^2(I/2)\cot^{\zeta-1}(I/2)$, which are independent of the disk models and mass of the captured objects, are found in the capture process for stars and BHs, respectively. These integrals of motion are useful in predicting the final captured orbital semi-major axis in AGN disks.

    \item The radial density profile of captured objects in AGN disks solely depends on the angular density and eccentricity distribution of the stars/BHs in nuclei clusters and is effectively independent of the radial density profile of the stars/BHs in nuclei clusters and disk models. An isotropic nuclei cluster with a thermal eccentricity distribution will predict $\dd N/\dd r\propto r^{-1/4}$ for captured objects in AGN disks.

    \item If the AGN disk is a self-gravitating disk with a surface density scaling as $\Sigma\propto r^{-3/2}$, {besides the initially embedded stars}, the total mass of the captured stars can be well described by a mass filling function $\mathcal{M}_*(t)=2M_{\rm SMBH}f_* \left(\frac{\Sigma_m}{\Sigma_*}\frac{t}{T_m}\right)^{1-\gamma_{\rm NSC}/3}$. The captured stars will be distributed between $r_g$ and $r_m\left(\frac{\Sigma_m}{\Sigma_*}\frac{t}{T_m}\right)^{1/3}$, with a number profile $\dd N/\dd r\propto r^{-1/4}$.

    \item If the AGN disk is a self-gravitating disk, {besides the initially embedded BHs}, the total mass of the captured BHs can be roughly described by a mass filling function $\mathcal{M}_{\rm BH}(t)=2M_{\rm SMBH}f_{\rm BH}\left(\frac{m}{M_{\rm SMBH}}\right)\left(\frac{\Sigma_m\pi r_m^2}{M_{\rm SMBH}}\right)\left(\frac{t}{T_m}\right)$. The captured BHs will be distributed between $r_g$ and $r_m$ with a number profile $\dd N/\dd r\propto r^{-1/4}$.

    \item Captured stars/BHs could become dynamically crowded through disk capture. The AGN disk has the potential to capture and accumulate individual objects from nuclei clusters in the radial direction, which could lead to the direct formation of binary star/BH systems immediately after capture within the disk. In this scenario, the controversial migration traps may not be necessary for BH/star mergers to occur in AGN disks.
\end{itemize}

\section*{Acknowledgements}
YW is supported by Nevada Center for Astrophysics. Z.Z. acknowledges support from the National. Science Foundation under CAREER grant AST-1753168 and support from NASA award 80NSSC22K1413. 
%%%%%%%%%%%%%%%%%%%%%%%%%%%%%%%%%%%%%%%%%%%%%%%%%%
\section*{Data Availability}
The data underlying this article will be shared on reasonable request to the corresponding author.



\bibliographystyle{mnras}
\bibliography{example} % if your bibtex file is called example.bib

%\begin{thebibliography}{99}
%\end{thebibliography}

%%%%%%%%%%%%%%%%%%%%%%%%%%%%%%%%%%%%%%%%%%%%%%%%%%

%%%%%%%%%%%%%%%%% APPENDICES %%%%%%%%%%%%%%%%%%%%%

\appendix

%%%%%%%%%%%%%%%%%%%%%%%%%%%%%%%%%%%%%%%%%%%%%%%%%%

% Don't change these lines
\bsp	% typesetting comment
\label{lastpage}
\end{document}